\documentclass[journal,10pt,onecolumn]{IEEEtran}
\usepackage{amssymb,amsmath,graphicx}
\usepackage{algorithmicx}
\usepackage{algpseudocode}
\usepackage{mdframed}
\usepackage{xspace}
\usepackage{graphicx}
\usepackage{graphics}
\usepackage{caption}
\usepackage{subcaption}
\usepackage{subfig}
\usepackage{multirow}
\usepackage{epstopdf}
\usepackage{array}
\usepackage{cite}
\usepackage{enumerate}
\usepackage{lipsum}
\usepackage{color}

\usepackage{fancyhdr}

\makeatletter
\let\old@ps@headings\ps@headings
\let\old@ps@IEEEtitlepagestyle\ps@IEEEtitlepagestyle
\def\confheader#1{%
  \def\ps@headings{%
    \old@ps@headings%
    \def\@oddhead{\strut\hfill#1\hfill\strut}%
    \def\@evenhead{\strut\hfill#1\hfill\strut}%
  }%
  \def\ps@IEEEtitlepagestyle{%
    \old@ps@IEEEtitlepagestyle%
    \def\@oddhead{\strut\hfill#1\hfill\strut}%
    \def\@evenhead{\strut\hfill#1\hfill\strut}%
  }%
  \ps@headings%
}
\makeatother

\confheader{%
  Accepted for publication in IEEE Transactions on Aerospace and Electronics Systems, March 2016
}

\newcommand{\ie}{i.e.}

\ifCLASSINFOpdf
\else
\fi
\begin{document}
%
\title{Multisensor--Multitarget Bearing--Only Sensor Registration}
%
%
%
\author{\authorblockN{Ehsan Taghavi, R. Tharmarasa \\and T. Kirubarajan\\}
\authorblockA{McMaster University\\
Hamilton, Ontario, Canada\\
email: $\{${taghave, tharman, kiruba}$\}$@mcmaster.ca}\\
\and \authorblockN{Mike McDonald\\} \authorblockA{Defence Research and Development Canada\\
Ottawa, Ontario, Canada\\
email:  mike.mcdonald@drdc-rddc.gc.ca}}
\maketitle

\begin{abstract}
\textcolor{black}{Bearing--only estimation is one of the fundamental and challenging problems in target tracking. As in the case of radar tracking, the presence of offset or position biases can exacerbate the challenges in bearing--only estimation. Modeling various sensor biases is not a trivial task and not much has been done in the literature specifically for bearing--only tracking. This paper addresses the modeling of offset biases in bearing--only sensors and the ensuing multitarget tracking with bias compensation. Bias estimation is handled at the fusion node to which individual sensors report their local tracks in the form of associated measurement reports (AMR) or angle-only tracks. The modeling is based on a multisensor approach that can effectively handle a time--varying number of targets in the surveillance region. The proposed algorithm leads to a maximum likelihood bias estimator. The corresponding Cram\'er--Rao Lower Bound to quantify the theoretical accuracy that can be achieved by the proposed method or any other algorithm is also derived. Finally, simulation results on different distributed tracking scenarios are presented to demonstrate the capabilities of the proposed approach. In order to show that the proposed method can work even with false alarms and missed detections, simulation results on a centralized tracking scenario where the local sensors send all their measurements (not AMRs or local tracks) are also presented.}
\end{abstract}

\begin{IEEEkeywords}
Bias estimation, bearing--only tracking, target motion analysis, triangulation, filtering, multisensor--multitarget
\end{IEEEkeywords}

%
\IEEEpeerreviewmaketitle

\section{Introduction}
Multisensor--multitarget bearing--only tracking is a challenging problem with many applications \cite{bar2001estimation,bar2011tracking,kronhamn1998bearings,chan1992bearings}. Some applications of bearing--only tracking are in maritime surveillance using sonobuoys, underwater target tracking using sonar and passive ground target tracking using Electronic Support Measures (ESM) or Infra--red Search and Track (IRST) sensors. In such applications, it is of interest to find the target position as well as any biases that may affect estimation performance. From the early works in \cite{aidala1979kalman,aidala1983utilization,nardone1984fundamental} to the more recent works in \cite{leong2013gaussian} and the references therein, the focus has been only on tracking the targets based on measurements from a bearing--only sensor. However, due to the limitations of single sensor bearing--only tracking, \ie, due to the need for own--ship maneuvers for the observability of state parameters \cite{la2008analysis}, the issue of biases in passive single sensor tracking has not been addressed in the literature. The main focus of this paper is multisensor bearing--only tracking in the presence of biases. In multisensor bearing-only tracking, observability \textcolor{black}{is no longer a major issue}. \textcolor{black}{However, in the case of port--starboard ambiguity, the problem of observability was discussed in detail in \cite{braca2014bayesian}.} \textcolor{black}{Besides}, the presence of sensor biases that are often unaccounted for can degrade the estimation results significantly. Most of the works on bias estimation have been about radar tracking \textcolor{black}{(see \cite{taghavi2014bias, taghavi2013bias, fortunati2011least}} and the references therein) or using other measurements besides bearing information \cite{belfadel2013bias,belfadel2013minimalist}. For example, when the elevation information is available, one can estimate the offset biases as in \cite{belfadel2013minimalist,chen2012bias}.

With the objective of providing a combined bias estimation and target tracking algorithm that is both theoretically sound and practical, the problem of multisensor bearing-only multitarget tracking is considered in this paper. Having more than one passive sensor in the surveillance region ensures the observability of the state parameters, \ie, position and velocity of the target, without the need for maneuvers \cite{mallick2001multi,xu2007biased,muvsicki2008bearings}. One of the issues that can complicate bearing--only tracking is the bias in the sensors. For example, in maritime surveillance using sonobuoys, which are usually dropped from an aircraft or thrown from a ship close to an area of interest, the exact locations of the sonobuoys are not known. This leads to position biases \cite{johansson1997submarine}. \textcolor{black}{This is also an isuue in modern systems such as autonomous underwater vehicles (AUV) \cite{bracadistributed}.}  In addition, the impact with the water surface and the waves can result in systemic offset biases \cite{belfadel2013bias}. In wide area surveillance using airborne IRST sensors, uncertain platform motion can contribute to biases as well. Offset bias can be modeled as an additive constant term affecting the measurement equation and the sensor position uncertainty can be modeled using a random walk \cite{lin2004exact}.

Negligible biases can be treated as residual errors. This residual error can be used in the form of additional uncertainty in the measurements later in the filtering step. However, if offset biases are larger than the noise standard deviation of the bearing--only measurements, a mirror of the target's bearing is sent to fusion center instead of its actual value, which will result in totally erroneous estimates. Further, these erroneous measurements can worsen state estimation results when fused with measurements from similarly biased sensors. In order to benefit from the information available from multiple bearing--only sensors with offset bias, one needs to model, estimate and then correct the biases. This is precisely the motivation for this paper. 

The focus of this paper is offset bias estimation. In order to model offset biases, one can transform the measurement space of sensor data to Cartesian coordinates followed by covariance matrix transformation. This transformation will make it possible to find an exact model for the biases that can be used in bias estimation and correction. However, the full position information is not available in a single bearing--only measurement. The process of measurement transformation is done by paring measurements from different sensors in the surveillance region. That is, this transformation is done through triangulation \cite{hendeby2006recursive}. With the new pseudo--measurement in Cartesian coordinates, position and velocity of the targets can be estimated over time as new measurements are generated from paired sensors at subsequent times \cite{muvsicki2008bearings,farina1999target,arulampalam2004bearings,iltis1996consistent}. Assuming that these state estimates also carry the effects of offset biases, it is possible to find such biases, if any, and correct them. In the case of bistatic passive sensors, one can use the methods in \cite{xu2007biased,okello2003maximum}. However, previous work on multisensor bearing--only bias estimation is still limited. The method presented here gives a comprehensive analysis of offset bias modeling in multisensor passive bearing--only sensors.

The proposed method gives an exact model for bearing--only biases in Cartesian coordinates. In addition, the formulation of an appropriate likelihood function enables the use of maximum likelihood estimators to find the biases. Also, the proposed model is robust against large sensor noise standard deviation. Finally, as shown through simulation results, large bearing biases can be estimated accurately, which leads to correspondingly accurate target state estimation results.

The goal of this paper is to present a step--by--step approach for designing a multisensor--multitarget tracking system based on biased bearing--only measurements and give a practical solution to the problem of bias estimation. In Section \ref{paperB_formulation}, a detailed model of the multisensor bearing--only estimation problem with bias is given. Section \ref{paperB_BiasModel} is devoted to modeling the offset biases in Cartesian coordinates. In Section \ref{paperB_BOT}, a practical solution for bias estimation is proposed. Section \ref{paperB_CRLB_bearing} presents the derivation of Cram\'er--Rao lower bounds. Simulation results are shown in Section \ref{paperB_results} along with discussions on different scenarios. Finally, Section \ref{paperB_conclusion} ends the paper with conclusions.

\section{\label{paperB_formulation} Bearing--Only Estimation Problem}
\begin{figure}[htbp!]
\centering
\includegraphics[width=3.5in]{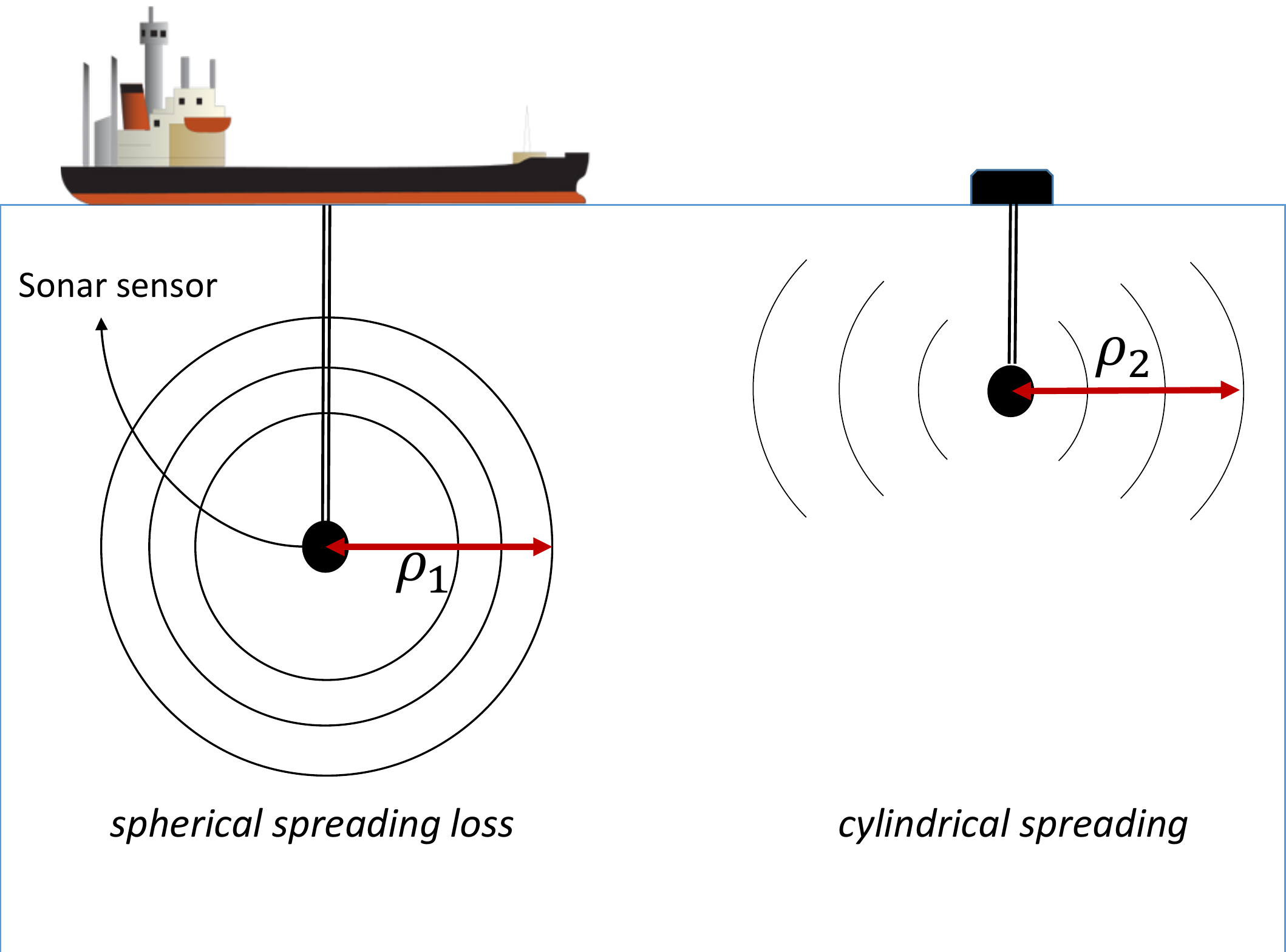}
\caption{Bearing--only sensors and signal propagation types \cite{kolev2011sonar}}
\label{paperB_fig:ships}
\end{figure}

Bearing--only sensors with operating ranges of hundreds of meters to a few kilometers are one of the most crucial sensors in maritime or ground surveillance applications. These sensors can actively or passively detect the directions of arrival of signals emitted by the targets of interest. While underwater surveillance is the common application of bearing--only tracking, it is also used in surface and air target tracking. For example, ESM and IRST sensors also use bearing--only sensors for tracking. As shown in Figure \ref{paperB_fig:ships}, bearing--only sensors can be on the own--ship or deployed separately in the surveillance region. Moreover, they can operate under different environmental conditions as shown in Figure \ref{paperB_fig:ships} \cite{blank2005introduction,kolev2011sonar}.

The bearing--only measurement from passive sensors is written as
\begin{eqnarray}
z_{s}(k) & = & \theta_{s}(k)+b_{s}+w_{s}(k),
\label{paperB_eq:BO_measurement}
\end{eqnarray}
where $z_{s}(k)$ is the direction of arrival at sensor $s$, $\theta_{s}$ is the true bearing of the target, $b_{s}$ is the constant bias in the measurements of sensor $s$ and $w_{s}$ is an additive zero--mean white Gaussian noise with variance $\sigma^2_{\theta}$. It is assumed that there are $S$ bearing--only sensors in the surveillance region at positions $(x_s,y_s)$ for $s=1,2,...,S$, and, they record targets' bearings at time instants $k=1,...,K$. Note that there is no index to denote target ID, but wherever such clarification is needed, it will be included.

\textcolor{black}{In this paper, bias estimation is handled at the fusion node to which individual sensors report their local tracks in the form of associated measurement reports (AMR) \cite{yosinski2009network,cooperman2002tactical} or angle-only tracks. That is, a distributed tracking system is considered as in the case of \cite{lin2004exact,lin2005multisensor,taghavi2014bias}. However, the difference is that in these earlier works local full-state tracks were sent to the fusion node for bias estimation whereas AMRs or bearing-only tracks are sent in the present case where the local sensors do not have full observability due to bearing-only measurements. As in \cite{lin2004exact,lin2005multisensor,taghavi2014bias}, working with tracks or AMRs obviates the need at the fusion center to address false alarms and missed detections that are inevitable at the local sensors although ghost tracks may be present among local tracks. However, in order to show that the proposed method can work even with false alarms and missed detections in a centralized tracking system, in one experiment in Section \ref{paperB_results} we assume that the local sensors send all their measurements (including false alarms and missed detections instead of AMRs or local bearing-only tracks) and evaluate the performance of the proposed method. In \cite{okello2003maximum} and \cite{xu2007biased}, where bias estimation at measurement level (rather than at track level as in \cite{lin2004exact,lin2005multisensor,taghavi2014bias}) is considered, false alarms and missed detections are not addressed at all. In \cite{li2010joint}, a joint data association and bias estimation method was proposed for linear measurement models, which is not applicable for bearing-only systems. A general case of multistatic passive radar system with false alarms and missed detections was considered in \cite{daun2012tracking} and \cite{choi2014approaches}, but, the bias problem was not addressed.}

The goal of bearing--only tracking is to find the bias in each sensor and estimate each target's position as accurately as possible based on the model given in \eqref{paperB_eq:BO_measurement}, either as decoupled parameters or jointly \cite{bugallo2007bearings,xu2007cramer}. Due to the computational burdens of joint tracking and parameter estimation methods \cite{hopkins1988pseudo}, a decoupled bias and state estimation is presented in this paper. 

It is assumed that each target is following the Discretized Continuous White Noise Acceleration (DCWNA) or the nearly constant velocity (NCV) model \cite{bar2011tracking}. As a result, a target's state vector in 2D Cartesian coordinate is given by
\begin{eqnarray}
\mathbf{x}\left(k\right) & = & \left[\begin{array}{cccc}
x(k) & v_{x}(k) & y(k) & v_{y}(k)\end{array}\right]^{\mathrm{T}}\label{paperB_eq:vectorX}
\end{eqnarray}
with $\left(x(k),y(k)\right)$ being the position and $\left(v_{x}(k),v_{y}(k)\right)$ being the velocity. The motion model can further be defined as
\begin{eqnarray}
\mathbf{x}\left(k+1\right) & = & F(k)\mathbf{x}\left(k\right)+\nu(k)\label{paperB_eq:targetMotion}
\end{eqnarray}
where the state transition matrix is
\begin{eqnarray}
F(k) & \triangleq & \left[\begin{array}{cccc}
1 & T & 0 & 0\\
0 & 1 & 0 & 0\\
0 & 0 & 1 & T\\
0 & 0 & 0 & 1
\end{array}\right]\label{paperB_eq:Fdefine}
\end{eqnarray}
and the covariance matrix of $\nu(k)$ which is an additive zero--mean white Gaussian noise vector is
\begin{eqnarray}
Q & = & \left[\begin{array}{cc}
Q_{x} & 0\\
0 & Q_{y}
\end{array}\right]\label{paperB_eq:Qdefine}
\end{eqnarray}
in which $Q_{x}$ and $Q_{y}$ are defined as
\begin{eqnarray}
Q_{x} & = & \left[\begin{array}{cc}
\frac{1}{3}T^{3} & \frac{1}{2}T^{2}\\
\frac{1}{2}T^{2} & T
\end{array}\right]\bar{q}_{x}\label{paperB_eq:Qxdefine}\\
Q_{y} & = & \left[\begin{array}{cc}
\frac{1}{3}T^{3} & \frac{1}{2}T^{2}\\
\frac{1}{2}T^{2} & T
\end{array}\right]\bar{q}_{y}\label{paperB_eq:Qydefine}
\end{eqnarray}
where $\bar{q}_{x}$ and $\bar{q}_{y}$ are noise intensities with dimension ${\mathrm{m}^{2}}/{\mathrm{s}^{3}}$ \cite{bar2001estimation}.

In order to transform the measurements from polar to Cartesian coordinates, it is initially assumed that $S$ is even and that the sensors are paired into $\frac{S}{2}$ one--to--one pairs. Note that this constraint is relaxed in Section \ref{paperB_BOT}. For a pair of sensors with a single target in the common field of view, the best estimate for the location of the target, independent of its previous location, can be obtained through triangulation \cite{chen2012bias,qi2007modified,hendeby2006recursive}. The triangulated estimates of the target position at time $k$ using sensor pair $(i,j)$, ignoring measurement noise, are given by
\begin{eqnarray}
\hat{x}_{ij}(k) & = & \frac{y_{j}-y_{i}+x_{i}\tan\left(\theta_{i}(k)+b_{i}\right)}{\tan\left(\theta_{i}(k)+b_{i}\right)-\tan\left(\theta_{j}(k)+b_{j}\right)}
-\frac{x_{j}\tan\left(\theta_{j}(k)+b_{j}\right)}{\tan\left(\theta_{i}(k)+b_{i}\right)-\tan\left(\theta_{j}(k)+b_{j}\right)}\nonumber \\
\label{paperB_eq:triangulation_x}
\end{eqnarray}

\begin{eqnarray}
\hat{y}_{ij}(k) & = & \frac{y_{j}\tan\left(\theta_{i}(k)+b_{i}\right)-y_{i}\tan\left(\theta_{j}(k)+b_{j}\right)}{\tan\left(\theta_{i}(k)+b_{i}\right)-\tan\left(\theta_{j}(k)+b_{j}\right)}
+\frac{(x_{i}-x_{j})\tan\left(\theta_{i}(k)+b_{i}\right)\tan\left(\theta_{j}(k)+b_{j}\right)}{\tan\left(\theta_{i}(k)+b_{i}\right)-\tan\left(\theta_{j}(k)+b_{j}\right)}\nonumber \\
\label{paperB_eq:triangulation_y}
\end{eqnarray}
where $\hat{x}_{ij}(k)$ and $\hat{y}_{ij}(k)$ are the $X$ and $Y$ Cartesian estimates, respectively.

In addition, by defining the stacked covariance matrix in the bearing--only coordinate for the stacked measurement $\left[\begin{array}{cc}
\theta_{i}(k) & \theta_{j}(k)\end{array}\right]^{\mathrm{T}}$ as
\begin{eqnarray}
R_{ij}(k) & = & \left[\begin{array}{cc}
\sigma_{\theta_{i}(k)}^{2} & 0\\
0 & \sigma_{\theta_{j}(k)}^{2}
\end{array}\right]\label{paperB_eq:covariance_BO}
\end{eqnarray}
one can calculate the covariance matrix of the transformed vector $\left[\begin{array}{cc}
x_{ij}(k) & y_{ij}(k)\end{array}\right]^{\mathrm{T}}$ as
\begin{eqnarray}
R_{ij}^{xy}(k) & = & \left(J(k)^{\mathrm{T}}\left(R_{ij}(k)\right)^{-1}J(k)\right)^{-1}\label{paperB_eq:covariance_xy}
\end{eqnarray}
where $J(k)$ is the Jacobian matrix with respect to $\left[\begin{array}{cc}
x_{ij}(k) & y_{ij}(k)\end{array}\right]^{\mathrm{T}}$  and
\begin{eqnarray}
J(k) & = & \left[\begin{array}{cc}
J_{i}(k) & J_{j}(k)\end{array}\right]^{\mathrm{T}}\label{paperB_eq:Jfunc_ij}
\end{eqnarray}
Further, the elements of $J(k)$ can be written as
\begin{eqnarray}
J_{i}(k) & = & \left[\frac{-\left(y_{ij}(k)-y_{i}\right)}{\left(x_{ij}(k)-x_{i}\right)^{2}+\left(y_{ij}(k)-y_{i}\right)^{2}}\right.,
 \left.\frac{\left(x_{ij}(k)-x_{i}\right)}{\left(x_{ij}(k)-x_{i}\right)^{2}+\left(y_{ij}(k)-y_{i}\right)^{2}}\right]\label{paperB_eq:Jfunc_Ji}
\end{eqnarray}
and
\begin{eqnarray}
J_{j}(k) & = & \left[\frac{-\left(y_{ij}(k)-y_{j}\right)}{\left(x_{ij}(k)-x_{j}\right)^{2}+\left(y_{ij}(k)-y_{j}\right)^{2}}\right.,
\left.\frac{\left(x_{ij}(k)-x_{j}\right)}{\left(x_{ij}(k)-x_{j}\right)^{2}+\left(y_{ij}(k)-y_{j}\right)^{2}}\right]\label{paperB_eq:Jfunc_Jj}
\end{eqnarray}
If no bias estimation is needed, these pseudo--measurements in Cartesian coordinates can be used in a Kalman filter  with their associated covariance matrices to recursively estimate the target's position \cite{catlin2012estimation}.

\section{\label{paperB_BiasModel}Bias Modeling in Cartesian Coordinates}
As discussed in Section \ref{paperB_formulation}, bearing--only biases are in the form of the additive constants. Although additive constant biases have already been dealt with in the case of radar measurements \cite{taghavi2014bias} and \cite{taghavi2013bias}, it is not possible to generate similar pseudo--measurements with bearing--only data directly. One way to formulate pseudo--measurements with biases is to model in Cartesian coordinates. In this section, a step--by--step procedure to model the biases and to separate them from original track positions in Cartesian coordinates is given. In Section \ref{paperB_BOT}, the pseudo--measurement generation is discussed in detail.

In Section \ref{paperB_formulation}, the process of mapping from bearing--only measurements to Cartesian was given. In order to model the biases, one can start with separating the bias terms in \eqref{paperB_eq:triangulation_x} and \eqref{paperB_eq:triangulation_y} from the original track position in Cartesian coordinates. This separation of bias terms provides the necessary information to create a pseudo--measurement that properly addresses the biases as in the case of radar measurements. Once the pseudo--measurements are generated, it is possible to estimate the biases and remove them. The process of finding the bias terms that contribute to \eqref{paperB_eq:triangulation_x} and \eqref{paperB_eq:triangulation_y} starts with expanding the $\tan\left(\cdot\right)$ function as
\begin{eqnarray}
\tan\left(\alpha+\beta\right) & = & \frac{\tan\left(\alpha\right)+\tan\left(\beta\right)}{1-\tan\left(\alpha\right)\tan\left(\beta\right)}\label{paperB_eq:tangent}
\end{eqnarray}
Applying \eqref{paperB_eq:tangent} to \eqref{paperB_eq:triangulation_x} and \eqref{paperB_eq:triangulation_y}, and separating the terms related to the bias from those related to the target position will give a new set of equations to define the position of the target in Cartesian coordinates. To make the parameter separation easier to follow, the common terms are defined and named first. For the common terms in $\hat{x}_{ij}(k)$ and $\hat{y}_{ij}(k)$, one can define
\begin{eqnarray}
D & = & \tan\left(\theta_{i}(k)\right)-\tan\left(\theta_{j}(k)\right)\label{paperB_eq:tan_D}\\
B & = & 1+\tan\left(\theta_{i}(k)\right)\tan\left(\theta_{j}(k)\right)\label{paperB_eq:tan_B}
\end{eqnarray}
Further, define the following for the terms in $\hat{x}_{ij}(k)$:
\begin{eqnarray}
D_{x} & = & x_{i}\tan\left(\theta_{i}(k)\right)-x_{j}\tan\left(\theta_{j}(k)\right)+\left(y_{j}-y_{i}\right)\label{paperB_eq:tan_Dx}\\
B_{x}^{i} & = & -\left(y_{j}-y_{i}\right)\tan\left(\theta_{i}(k)\right)
+x_{j}\tan\left(\theta_{i}(k)\right)\tan\left(\theta_{j}(k)\right)+x_{i}\label{paperB_eq:tan_Bxi}\\
B_{x}^{j} & = & -\left(y_{j}-y_{i}\right)\tan\left(\theta_{j}(k)\right)
-x_{i}\tan\left(\theta_{i}(k)\right)\tan\left(\theta_{j}(k)\right)+x_{j}\label{paperB_eq:tan_Bxj}\\
B_{x}^{ij} & = & x_{j}\tan\left(\theta_{i}(k)\right)-x_{i}\tan\left(\theta_{j}(k)\right)
+\left(y_{j}-y_{i}\right)\tan\left(\theta_{i}(k)\right)\tan\left(\theta_{j}(k)\right)\label{paperB_eq:tan_Bxij}
\end{eqnarray}
Similarly, for the terms in $\hat{y}_{ij}(k)$, define
\begin{eqnarray}
D_{y} & = & y_{j}\tan\left(\theta_{i}(k)\right)-y_{i}\tan\left(\theta_{j}(k)\right)
+\left(x_{j}-x_{i}\right)\tan\left(\theta_{i}(k)\right)\tan\left(\theta_{j}(k)\right)\label{paperB_eq:tan_Dy}\\
B_{y}^{i} & = & \left(x_{j}-x_{i}\right)\tan\left(\theta_{j}(k)\right)
+y_{i}\tan\left(\theta_{i}(k)\right)\tan\left(\theta_{j}(k)\right)+y_{j}\label{paperB_eq:tan_Byi}\\
B_{y}^{j} & = & \left(x_{j}-x_{i}\right)\tan\left(\theta_{i}(k)\right)
-y_{j}\tan\left(\theta_{i}(k)\right)\tan\left(\theta_{j}(k)\right)-y_{i}\label{paperB_eq:tan_Byj}\\
B_{y}^{ij} & = & y_{i}\tan\left(\theta_{i}(k)\right)-y_{j}\tan\left(\theta_{j}(k)\right)
+\left(x_{j}-x_{i}\right)\label{paperB_eq:tan_Byij}
\end{eqnarray}
With these factorizations, bias terms can be separated from the target state values in Cartesian coordinates. It can be seen that the vector $\left[\begin{array}{cc}
\hat{x}_{ij}(k) & \hat{y}_{ij}(k)\end{array}\right]^{\mathrm{T}}$ can be written as\footnote{The superscript ``u'' indicates that the parameter is unbiased.}
\begin{eqnarray}
\left[\begin{array}{c}
 \hat{x}_{ij}(k)\\
\hat{y}_{ij}(k)
\end{array}\right] & = & \left[\begin{array}{c}
x_{ij}^{\mathrm{u}}(k)+\beta_{x}\left(\theta_{i}(k),\theta_{j}(k),b_{i},b_{j}\right)\\
y_{ij}^{\mathrm{u}}(k)+\beta_{y}\left(\theta_{i}(k),\theta_{j}(k),b_{i},b_{j}\right)
\end{array}\right]\nonumber \\
\label{paperB_eq:biasModelVector}
\end{eqnarray}
where
\begin{eqnarray}
x_{ij}^{\mathrm{u}}(k) & = & \frac{D_{x}}{D}\label{paperB_eq:xFree}\\
y_{ij}^{\mathrm{u}}(k) & = & \frac{D_{y}}{D}\label{paperB_eq:yFree}
\end{eqnarray}
and the bias terms can be written in as
\begin{eqnarray}
\beta_{x}\left(\theta_{i}(k),\theta_{j}(k),b_{i},b_{j}\right) & = & \frac{B_{x}^{i}\tan\left(b_{i}\right)+B_{x}^{j}\tan\left(b_{j}\right)+B_{x}^{ij}\tan\left(b_{i}\right)\tan\left(b_{j}\right)}{D+D\tan\left(b_{i}\right)+B\tan\left(b_{j}\right)+D\tan\left(b_{i}\right)\tan\left(b_{j}\right)}\nonumber \\
 &  & -\frac{D_{x}B\tan\left(b_{i}\right)-D_{x}B\tan\left(b_{j}\right)+D_{x}D\tan\left(b_{i}\right)\tan\left(b_{j}\right)}{D^{2}+BD\tan\left(b_{i}\right)-BD\tan\left(b_{j}\right)+D^{2}\tan\left(b_{i}\right)\tan\left(b_{j}\right)}\label{paperB_eq:beta_x}
\end{eqnarray}

\begin{eqnarray}
\beta_{y}\left(\theta_{i}(k),\theta_{j}(k),b_{i},b_{j}\right) & = & \frac{B_{y}^{i}\tan\left(b_{i}\right)+B_{y}^{j}\tan\left(b_{j}\right)+B_{y}^{ij}\tan\left(b_{i}\right)\tan\left(b_{j}\right)}{D+D\tan\left(b_{i}\right)+B\tan\left(b_{j}\right)+D\tan\left(b_{i}\right)\tan\left(b_{j}\right)}\nonumber \\
 &  & -\frac{D_{y}B\tan\left(b_{i}\right)-D_{y}B\tan\left(b_{j}\right)+D_{y}D\tan\left(b_{i}\right)\tan\left(b_{j}\right)}{D^{2}+BD\tan\left(b_{i}\right)-BD\tan\left(b_{j}\right)+D^{2}\tan\left(b_{i}\right)\tan\left(b_{j}\right)}\label{paperB_eq:beta_y}
\end{eqnarray}
Note that in the above formulations it is assumed that the biases and the true bearings are available. In practice, only the noisy or estimated values are available. Assuming small values for the bias and noise terms, one can use the above formulation without significant loss in accuracy. Similar assumptions has been made in the previous works on bias estimation \cite{taghavi2014bias,lin2004exact,lin2005multisensor}. A technical discussion on the range of bias and noise values for which the above formulation is valid is given in Section \ref{paperB_results}.

\section{\label{paperB_BOT}Bearing--Only Tracking and Registration}
Bearing--only sensor registration is a challenging problem in target tracking that has been addressed in \cite{song2010real,mcmichael1996maximum,okello2003maximum}. In order to find the biases and correct the measurements, one should first look into the observability of the bias variables. Note that in \eqref{paperB_eq:biasModelVector}, provided that the target is not on the line that connects the two sensors used in the triangulation or in the vicinity of one of the sensors, the state parameters are observable \cite{bar2001estimation}. In addition, if there are two pairs of sensors tracking the same target, the biases become observable as it is shown in \ref{paperB_pseduoBearing}. To estimate the biases decoupled from the state vector, a pseudo--measurement that can address the bias vector directly must be defined. In this section, a new formulation is proposed to create a pseudo--measurement that can be used for bias estimation with bearing--only data. The key requirement of this method in order to ensure observability of all parameters is to have at least two sensor pairs in the surveillance region. In the following, two practical scenarios that can be expanded to a more general formulation to handle varying number of sensors and targets are discussed in detail.

\subsection{\label{paperB_pseduoBearing}Pseudo--measurement of bearing--only measurements}
To handle practical bearing--only scenarios, two different cases are analyzed here. In each case, a separate pseudo--measurement model is proposed along with its associated covariance matrix. The main idea is to use two different position approximations to create a pseudo--measurement as discussed below.
\subsubsection{\label{paperB_4sensor}Four--sensor (or any even number of sensors) case}
Assuming that there are two pairs of sensors in the surveillance region, a vector of nonlinear pseudo--measurements can be defined by subtracting the target positions based on the pairs $(i,j)$ and $(m,n)$ as
\begin{eqnarray}
z_{b}(k) & = & \left[\begin{array}{c}
 \hat{x}_{ij}(k)-\hat{x}_{mn}(k)\\
\hat{y}_{ij}(k)-\hat{y}_{mn}(k)
\end{array}\right]+w(k)\label{paperB_eq:pseudoBO}
\end{eqnarray}
where $w(k)$ is the additive zero--mean white Gaussian noise associated with the pseudo--measurement and its covariance matrix is defined as 
\begin{eqnarray}
R_{w}(k) & = & R_{ij}^{xy}(k)+R_{ij}^{xy}(k)\label{paperB_eq:covPseudoBO}
\end{eqnarray}
Using the fact that in the absence of bias and noise terms, measurements from any two sensors point to the same target location regardless of the sensor locations, the pseudo--measurement $z_{b}(k)$ can be written as
\begin{eqnarray}
z_{b}(k) & = & \left[\begin{array}{c}
\beta_{x}\left(b_{i},b_{j}\right)-\beta_{x}\left(b_{m},b_{n}\right)\\
\beta_{y}\left(b_{i},b_{j}\right)-\beta_{x}\left(b_{m},b_{n}\right)
\end{array}\right]+w(k)\label{paperB_eq:pseudoBO-1}
\end{eqnarray}
for two uncorrelated pairs of sensors. This can be applied to any even number of sensors. 

\subsubsection{\label{paperB_3sensor}Three--sensor (or any odd number of sensors) case}
In this case, one must create two position approximations from triangulation to be able to create a pseudo--measurement for biases. Since there are only three sensors, the possible pairs are $(i,j)$ and $(i,m)$. Then, the pseudo--measurement can be approximated by
\begin{eqnarray}
z_{b}(k) & = & \left[\begin{array}{c}
 \hat{x}_{ij}(k)-\hat{x}_{im}(k)\\
\hat{y}_{ij}(k)-\hat{y}_{im}(k)
\end{array}\right]+w(k)\label{paperB_eq:pseudoBO_3sensors}
\end{eqnarray}
where $w(k)$ is approximately additive zero--mean white Gaussian noise associated with the pseudo--measurement and its covariance matrix is defined as 
\begin{eqnarray}
R_{w}(k) & = & R_{ij}^{xy}(k)+R_{im}^{xy}(k)\label{paperB_eq:covPseudoBO}
\end{eqnarray}
Because of the correlation between the two tracks from three sensors in Cartesian coordinates, the noise is not white anymore and this formulation is only an approximation. 

As in the case for four sensors, the pseudo--measurement $z_{b}(k)$ can be written as
\begin{eqnarray}
z_{b}(k) & = & \left[\begin{array}{c}
\beta_{x}\left(b_{i},b_{j}\right)-\beta_{x}\left(b_{i},b_{m}\right)\\
\beta_{y}\left(b_{i},b_{j}\right)-\beta_{x}\left(b_{i},b_{m}\right)
\end{array}\right]+w(k)\label{paperB_eq:pseudoBO-2}
\end{eqnarray}
Note that for simplicity, the arguments $\theta_{i}(k)$, $\theta_{j}(k)$, $\theta_{m}(k)$ and $\theta_{n}(k)$ have been dropped from \eqref{paperB_eq:pseudoBO-1} and \eqref{paperB_eq:pseudoBO-2}. With an arbitrary odd number of sensors, one sensor need to be paired with two other, resulting in in a similar approximation.
As for the more general case of $n$ arbitrary sensors, methods similar to \cite{taghavi2014bias} can be adopted to handle the situation.

\subsection{\label{paperB_BML}Batch maximum--likelihood estimator}
To apply a batch estimator for bias estimation, one needs to form a likelihood function. Following \eqref{paperB_eq:covPseudoBO} and \eqref{paperB_eq:pseudoBO-1}, and assuming that the noise is white, zero--mean and Gaussian, the likelihood function of the bias parameters given two pairs of sensors is  
\begin{eqnarray}
p\left(z_{b}(k)\mid\mathbf{b}\right) & = & \frac{1}{2\pi\left|R_{w}(k)\right|^{-\frac{1}{2}}}\exp\left(-\frac{1}{2}\left[z_{b}(k)-h(\mathbf{b})^{\mathrm{T}}\right]\right.\nonumber \\
 &  & \left.\left(R_{w}(k)\right)^{-1}\left[z_{b}(k)-h(\mathbf{b})\right]\right)\label{paperB_eq:likelihood_k}
\end{eqnarray}
where the nonlinear function $h(\mathbf{b})$ of the bias vector is given by
\begin{eqnarray}
h(\mathbf{b}) & = & \left[\begin{array}{c}
\beta_{x}\left(b_{i},b_{j}\right)-\beta_{x}\left(b_{m},b_{n}\right)\\
\beta_{y}\left(b_{i},b_{j}\right)-\beta_{x}\left(b_{m},b_{n}\right)
\end{array}\right]\label{paperB_eq:likelihood_hb}
\end{eqnarray}
Assuming independence over time, one can write the likelihood function over $k=1,...,K$ as
\begin{eqnarray}
p\left(Z_{b}\mid\mathbf{b}\right) & = & \prod_{k=1}^{K}p\left(z_{b}(k)\mid\mathbf{b}\right)\label{paperB_eq:likelihoodAll}
\end{eqnarray}
where
\begin{eqnarray}
Z_{b} & = & \left\{ z_{b}(1),z_{b}(2),\ldots,z_{b}(K)\right\} \label{paperB_paperB_eq:likelihood_Zb}
\end{eqnarray}
Finally, the vector $\hat{\mathbf{b}}$  that maximizes the likelihood function can be written as
\begin{eqnarray}
\hat{\mathbf{b}} & = & \arg\;\max_{\mathbf{b}}\; p\left(Z_{b}\mid\mathbf{b}\right)\label{paperB_eq:maxLikelihood}
\end{eqnarray}
The above assumes that there is only one target, but it can be extended to the multitarget case using the stacked parameter estimator.

\section{\label{paperB_CRLB_bearing}Cram\'er--Rao Lower bound for Bearing only bias estimation}
This section is devoted to the calculation of the Cram\'er--Rao Lower Bound (CRLB) on the estimation accuracy of bias parameters by using the pseudo--measurements introduced in Subsection \ref{paperB_pseduoBearing}. Note that based on \eqref{paperB_eq:likelihood_hb}, the measurement equation for target $t$ at time $k$ is
\begin{eqnarray}
h_{\mathbf{b}}^{t}(k) & = & \left[\begin{array}{c}
\beta_{x}^{t}\left(b_{i},b_{j}\right)-\beta_{x}^{t}\left(b_{m},b_{n}\right)\\
\beta_{y}^{t}\left(b_{i},b_{j}\right)-\beta_{x}^{t}\left(b_{m},b_{n}\right)
\end{array}\right]+w^{t}(k)\label{paperB_eq:likelihood_hb-multitarget}
\end{eqnarray}
Assuming that there are $N$ targets in the surveillance region and that the bias parameters are constant over time $k=1,...K$, the stacked measurement vector can be written as
\begin{eqnarray}
\mathbf{Z} & = & \mathbf{h}(\mathbf{b})+\mathbf{w}\label{paperB_eq:Y_multisensor}
\end{eqnarray}
where
\begin{eqnarray}
\mathbf{Z} & = & \left[\begin{array}{ccccccc}
z_{b}^{1}(1)^{\mathrm{T}} & \cdots & z_{b}^{N}(1)^{\mathrm{T}} & \cdots & z_{b}^{1}(K)^{\mathrm{T}} & \cdots & z_{b}^{N}(K)^{\mathrm{T}}\end{array}\right]^{\mathrm{T}}\\
\mathbf{h}(\mathbf{b}) & = & \left[\begin{array}{ccccccc}
h_{\mathbf{b}}^{1}(1)^{\mathrm{T}} & \cdots & h_{\mathbf{b}}^{N}(1)^{\mathrm{T}} & \cdots & h_{\mathbf{b}}^{1}(K)^{\mathrm{T}} & \cdots & h_{\mathbf{b}}^{N}(K)^{\mathrm{T}}\end{array}\right]^{\mathrm{T}}\\
\mathbf{w} & = & \left[\begin{array}{ccccccc}
w^{1}(1)^{\mathrm{T}} & \cdots & w^{N}(1)^{\mathrm{T}} & \cdots & w^{1}(K)^{\mathrm{T}} & \cdots & w^{N}(K)^{\mathrm{T}}\end{array}\right]^{\mathrm{T}}\label{paperB_eq:Y_matrices}
\end{eqnarray}
Further, the covariance matrix of the noise vector $\mathbf{w}$ is
\begin{eqnarray}
R & = & \mathrm{diag}\left(\left[\begin{array}{ccccccc}
R^{1}(1) & \cdots & R^{N}(1) & \cdots & R^{1}(K) & \cdots & R^{N}(K)\end{array}\right]\right)\label{paperB_eq:Y_R_multitarget}
\end{eqnarray}
The covariance matrix of an unbiased estimator $\hat{\mathbf{b}}$ is bounded from below by \cite{bar2001estimation}
\begin{equation}
\mathbb{E}\left\{ \left(\hat{\mathbf{b}}-\mathbf{b}\right)\left(\hat{\mathbf{b}}-\mathbf{b}\right)^{\mathrm{T}}\right\} \geq\mathbf{J}^{-1}
\label{paperB_paperB_eq:bearing_lowerBound}
\end{equation}
where $\mathbf{J}$ is the Fisher Information Matrix (FIM) given by
\begin{eqnarray}
\mathbf{J} & = & \mathbb{E}\left\{ \left[\nabla_{\mathbf{b}}\log p(\mathbf{Y}\mid\mathbf{b})\right]\left[\nabla_{\mathbf{b}}\log p(\mathbf{Y}\mid\mathbf{b})\right]^{\mathrm{T}}\right\} \mid_{\mathbf{b}=\tilde{\mathbf{b}}}\nonumber \\
 & = & \mathbb{E}\left\{ \left[\nabla_{\mathbf{b}}\lambda\right]\left[\nabla_{\mathbf{b}}\lambda\right]^{\mathrm{T}}\right\} \mid_{\mathbf{b}=\tilde{\mathbf{b}}}\label{paperB_eq:bearing_J}
\end{eqnarray}
in which $\tilde{\mathbf{b}}$ is the true value of the bias vector $\mathbf{b}$, $p(\mathbf{Z}\mid\mathbf{b})$ is the likelihood function of $\mathbf{b}$, $\lambda=-\ln p(\mathbf{Y}\mid\mathbf{b})$ and $\nabla$ is gradient operator.
Based on the model for the stacked measurement vector in \eqref{paperB_eq:Y_multisensor}, one can define the Jacobian matrix of $h_{\mathbf{b}}^{t}(k)$ evaluated at the true value $\mathbf{b}$ \cite{ristic2004beyond} as
\begin{eqnarray}
\tilde{H}_{\mathbf{b}}^{t}(k) & = & \left[\bigtriangledown_{\mathbf{b}}h_{\mathbf{b}}^{t}(k)^{\mathrm{T}}\right]^{\mathrm{T}}\label{paperB_eq:bearing_Jacobian}
\end{eqnarray}
Then, defining
\begin{eqnarray}
\tilde{\mathbf{H}}_{\mathbf{b}} & = & \left[\begin{array}{ccccccc}
\tilde{H}_{\mathbf{b}}^{1}(1)^{\mathrm{T}} & \cdots & \tilde{H}_{\mathbf{b}}^{N}(1)^{\mathrm{T}} & \cdots & \tilde{H}_{\mathbf{b}}^{1}(K)^{\mathrm{T}} & \cdots & \tilde{H}_{\mathbf{b}}^{N}(K)^{\mathrm{T}}\end{array}\right]^{\mathrm{T}}\label{paperB_eq:bearing_JacoboanMatrix}
\end{eqnarray}
one can write
\begin{eqnarray}
J & = & \mathbb{E}\left\{ \tilde{\mathbf{H}}_{\mathbf{b}}^{\mathrm{T}}R^{-1}\tilde{\mathbf{H}}_{\mathbf{b}}\right\} \label{paperB_paperB_eq:bearing_JFinal}
\end{eqnarray}

\section{\label{paperB_results}Simulation Results}
To evaluate how the proposed method performs in practical scenarios, a series of simulations is presented in this section. The implementation details on bias estimation, filtering and fusion  are also discussed. Simulation results on different scenarios are given with discussions on the advantages and disadvantages of the proposed bias estimator.

\subsection{\label{paperB_motion} Motion models and measurement generation}
\begin{figure}[htbp!]
\centering
\includegraphics[width=3.5in]{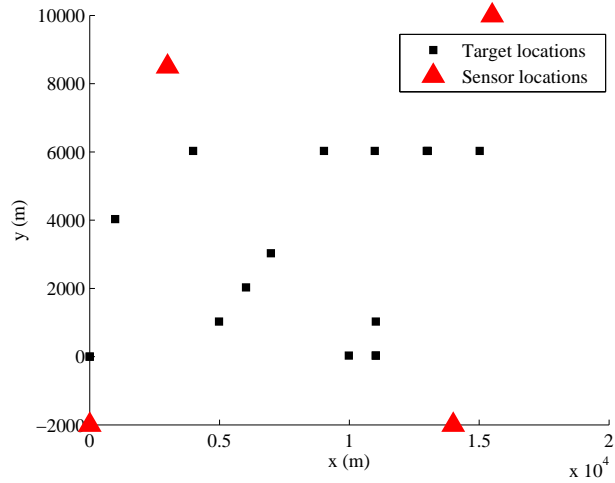}
\caption{Initial locations of the targets and sensors}
\label{fig:paperB_Scenario}
\end{figure}

A tracking scenario with four bearing--only sensors and sixteen targets is considered as shown in Figure \ref{fig:paperB_Scenario}. It is assumed that all sensors are synchronized and that the bias ranges are between $-0.05\mathrm{\: rad}$ and $0.05\mathrm{\: rad}$. The standard deviation of measurement noise is $\sigma_{\theta}=0.0261\;\mathrm{rad}=1.5^\circ$ for target bearing measurements, which is higher than what was previously assumed in the literature \cite{belfadel2013bias}.
The true motion of the targets is modeled using the DCWNA or the NCV model \cite{bar2001estimation} with $q_{x}=q_{y}=0.001{\mathrm{m}^{2}}/{\mathrm{s}^{3}}$. In the local trackers, DCWNA and Continuous Wiener Process Acceleration (CWPA) are used with $q_{x}=q_{y}=0.1{\mathrm{m}^{2}}/{\mathrm{s}^{3}}$ to create different scenarios for the simulation. In all scenarios, the sampling time is $T=1\mathrm{s}$.
To validate the proposed bias model and to quantify its performance at different bias values, the biases are set to both positive and negative values in different ranges as follows:
\begin{eqnarray}
\mathbf{b}_{\mathrm{test_{1}}} & = & \left[\begin{array}{cccc}
0.04\mathrm{\: rad} & -0.02\mathrm{\: rad} & 0.03\mathrm{\: rad} & -0.02\mathrm{\: rad}\end{array}\right]^{\mathrm{T}}\label{paperB_eq:test1}\\
\mathbf{b}_{\mathrm{test_{2}}} & = & \left[\begin{array}{cccc}
-0.04\mathrm{\: rad} & -0.02\mathrm{\: rad} & -0.03\mathrm{\: rad} & -0.02\mathrm{\: rad}\end{array}\right]^{\mathrm{T}}\label{paperB_eq:test2}\\
\mathbf{b}_{\mathrm{test_{3}}} & = & \left[\begin{array}{cccc}
0.04\mathrm{\: rad} & 0.02\mathrm{\: rad} & 0.03\mathrm{\: rad} & 0.02\mathrm{\: rad}\end{array}\right]^{\mathrm{T}}\label{paperB_eq:test3}
\end{eqnarray}

\subsection{\label{paperB_GA} Bias estimation and the Genetic Algorithm}
In this paper, the Genetic Algorithm (GA) \cite{davis1991handbook} is used to solve the optimization problem in \eqref{paperB_eq:maxLikelihood}. The Genetic Algorithm is an efficient optimization algorithm for highly nonlinear objective functions \cite{sharman1989genetic} that is widely used in different applications \cite{davis1991handbook}. Note that although the GA is a batch ML estimator, the length of the window can be varied depending on user criteria to meet the real--time requirements. The parameters used in the simulations are shown in Table \ref{paperB_GATable}.
\begin{table}
\caption{Parameter settings for the Genetic Algorithm}
\label{paperB_GATable}
\centering{}%
\begin{tabular}{|c|c|}
\hline 
Parameter & Value\tabularnewline
\hline 
\hline 
Lower bound & $-0.05\:\mathrm{rad}$\tabularnewline
\hline 
Upper bound & $0.05\:\mathrm{rad}$\tabularnewline
\hline 
Number of generations & $100$\tabularnewline
\hline 
Tolerance value & $1\times10^{-15}$\tabularnewline
\hline 
\end{tabular}
\end{table}
The algorithms were implemented on a computer with Intel\textsuperscript{\textregistered} Core\textsuperscript{\texttrademark} i7-3720Qm 2.60GHz processor and 8GB RAM.

\subsection{\label{paperB_comparision4sensors} Bias estimation: Four--sensor \textcolor{black}{distributed} problem}
In this scenario, all four sensors defined earlier are used to implement the GA. Four out of sixteen targets are used for performance evaluation. \textcolor{black}{AMRs or local bearing--only tracks} are collected over $100$ time steps and the GA is applied to the whole data in batch mode. The GA is run with the settings in Table \ref{paperB_GATable} and the final results are gathered after the termination of the GA. Then, the estimated bias vector is used over $100$ Monte Carlo runs to calculate the Root Mean Square Error (RMSE) for comparison. As the benchmark, the CRLB is also calculated based on the derivations in Section \ref{paperB_CRLB_bearing}. The RMSE values and $\sqrt{\mathrm{CRLB}\left\{ \left[\mathbf{b}\right]_{i}\right\}}$ of the ML estimates with the three different sets of bias parameters are shown in Tables \ref{tab:4sensors_tests1}, \ref{tab:4sensors_tests2} and \ref{tab:4sensors_tests3}.

\begin{table}[Htbp!]
\caption{Comparison of $\sqrt{\mathrm{CRLB}\left\{ \left[\mathbf{b}\right]_{i}\right\}}$ and RMSE of GA output for bias estimation of $\mathbf{b}_{\mathrm{test_{1}}}$ \textcolor{black}{in a distributed} system}
\label{tab:4sensors_tests1}
\begin{centering}
\begin{tabular}{|c|c|c|}
\hline 
Bias parameter & $\sqrt{\mathrm{CRLB}\left\{ \left[\mathbf{b}\right]_{i}\right\}}\:(\mathrm{rad})$ & RMSE$\:(\mathrm{rad})$ of the GA bias estimate\tabularnewline
\hline 
\hline 
$b_{i}$ & $0.071\times10^{-3}$ & $3.32\times10^{-3}$\tabularnewline
\hline 
$b_{j}$ & $0.071\times10^{-3}$ & $3.28\times10^{-3}$\tabularnewline
\hline 
$b_{m}$ & $0.537\times10^{-3}$ & $3.92\times10^{-3}$\tabularnewline
\hline 
$b_{n}$ & $0.462\times10^{-3}$ & $1.45\times10^{-3}$\tabularnewline
\hline 
\end{tabular}
\par\end{centering}
\end{table}

\begin{table}[Htbp!]
\caption{Comparison of $\sqrt{\mathrm{CRLB}\left\{ \left[\mathbf{b}\right]_{i}\right\}}$ and RMSE of the GA output for bias estimation of $\mathbf{b}_{\mathrm{test_{2}}}$ \textcolor{black}{in a distributed} system}
\label{tab:4sensors_tests2}
\centering{}%
\begin{tabular}{|c|c|c|}
\hline 
Bias parameter & $\sqrt{\mathrm{CRLB}\left\{ \left[\mathbf{b}\right]_{i}\right\}}\:(\mathrm{rad})$ & RMSE$\:(\mathrm{rad})$ of the GA bias estimate\tabularnewline
\hline 
\hline 
$b_{i}$ & $0.0904\times10^{-3}$ & $1.72\times10^{-3}$\tabularnewline
\hline 
$b_{j}$ & $0.0941\times10^{-3}$ & $1.81\times10^{-3}$\tabularnewline
\hline 
$b_{m}$ & $0.4243\times10^{-3}$ & $2.95\times10^{-3}$\tabularnewline
\hline 
$b_{n}$ & $0.3787\times10^{-3}$ & $1.73\times10^{-3}$\tabularnewline
\hline 
\end{tabular}
\end{table}

\begin{table}[Htbp!]
\caption{Comparison of $\sqrt{\mathrm{CRLB}\left\{ \left[\mathbf{b}\right]_{i}\right\}}$ and RMSE of GA output for bias estimation of $\mathbf{b}_{\mathrm{test_{3}}}$ \textcolor{black}{in a distributed} system}
\label{tab:4sensors_tests3}
\centering{}%
\begin{tabular}{|c|c|c|}
\hline 
Bias parameter & $\sqrt{\mathrm{CRLB}\left\{ \left[\mathbf{b}\right]_{i}\right\}}\:(\mathrm{rad})$ & RMSE$\:(\mathrm{rad})$ of the GA bias estimate\tabularnewline
\hline 
\hline 
$b_{i}$ & $0.0973\times10^{-3}$ & $2.73\times10^{-3}$\tabularnewline
\hline 
$b_{j}$ & $0.0954\times10^{-3}$ & $3.75\times10^{-3}$\tabularnewline
\hline 
$b_{m}$ & $0.5347\times10^{-3}$ & $2.43\times10^{-3}$\tabularnewline
\hline 
$b_{n}$ & $0.4880\times10^{-3}$ & $2.66\times10^{-3}$\tabularnewline
\hline 
\end{tabular}
\end{table}

Although there is a difference between the RMSE and $\sqrt{\mathrm{CRLB}\left\{ \left[\mathbf{b}\right]_{i}\right\}}$, the RMSE results are nearly an order of magnitude smaller than the bias values, which indicates that any correction made based on the estimated biases will result in better position estimates. Note that the CRLB in bearing--only tracking problems can be overly optimistic and may even approach zero (\ie, perfect estimates) in a network of bearing--only sensors \cite{horridge2003performance}. Thus, the difference between the theoretical CRLB and the empirical RMSE is not of major concern.

\begin{figure}[htbp!]
\centering
\includegraphics[width=3.5in]{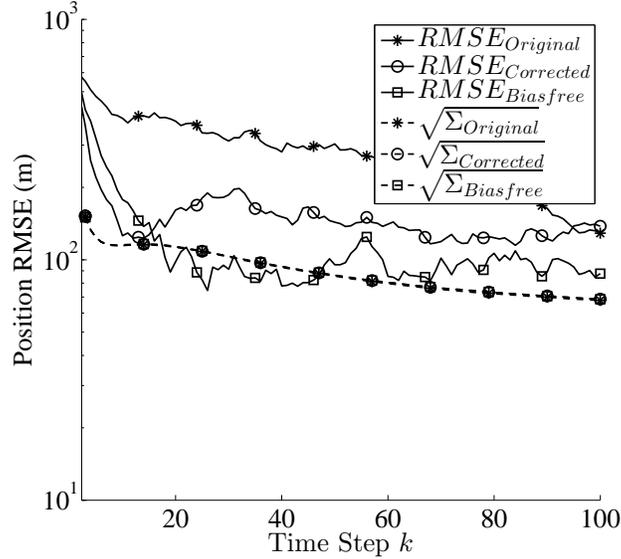}
\caption{Position RMSE \textcolor{black}{with distributed tracking} for corrected and original tracks of target 2 (set 1)}
\label{fig:paperB_Pos_set1_t2}
\end{figure}

\begin{figure}[htbp!]
\centering
\includegraphics[width=3.5in]{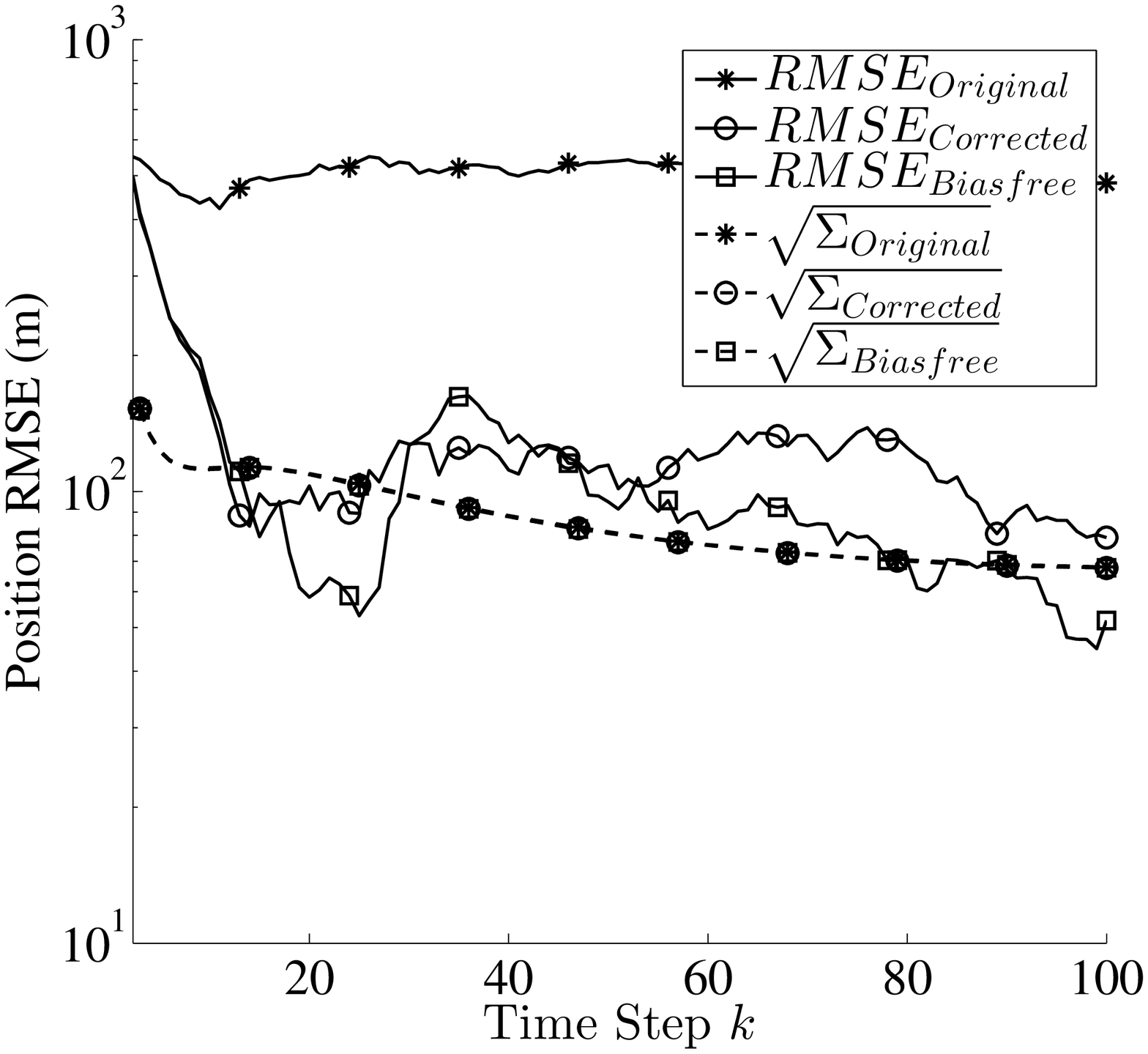}
\caption{Position RMSE \textcolor{black}{with distributed tracking} for corrected and original tracks of target 3 (set 1)}
\label{fig:paperB_Pos_set1_t3}
\end{figure}

To show how much the proposed bias estimation method can help in correcting the target tracks, another simulation is conducted. In this simulation, it is assumed that the tracker has access to the final estimated bias vector (output of the GA) and then a Kalman filter is run with the bias estimates in hand. To use the estimated bias parameters, one should, first, correct the bearing--only measurements with the estimated values. The correction must be done both in the measurement vector and its associated covariance matrix. Since the estimated biases do not have the covariance information, a scaled version of the the calculated CRLB of the bias parameters is used instead. The scaling factor can be determined through experiments. Then, the tracker can be run with these corrected measurements to find the position and velocity estimates of all targets in the surveillance region. The position RMSE of the original tracks before correction, the RMSE of the corrected estimates and the Cram\'er--Rao lower bounds are shown in Figures \ref{fig:paperB_Pos_set1_t2} and \ref{fig:paperB_Pos_set1_t3}. 

\begin{figure}[htbp!]
\centering
\includegraphics[width=3.5in]{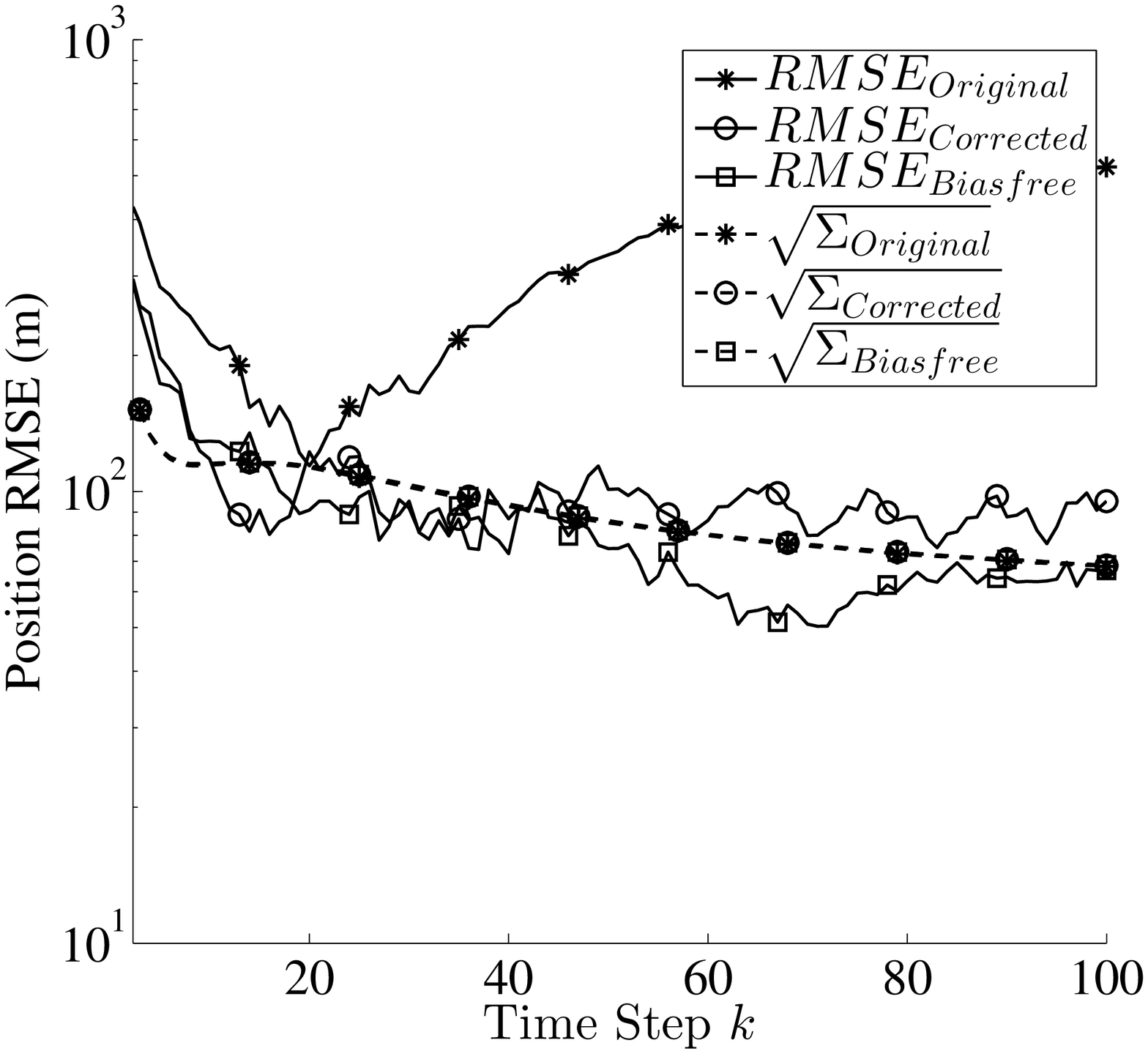}
\caption{Position RMSE \textcolor{black}{with distributed tracking} for corrected and original tracks of target 2 (set 2)}
\label{fig:paperB_Pos_set2_t2}
\end{figure}

\begin{figure}[htbp!]
\centering
\includegraphics[width=3.5in]{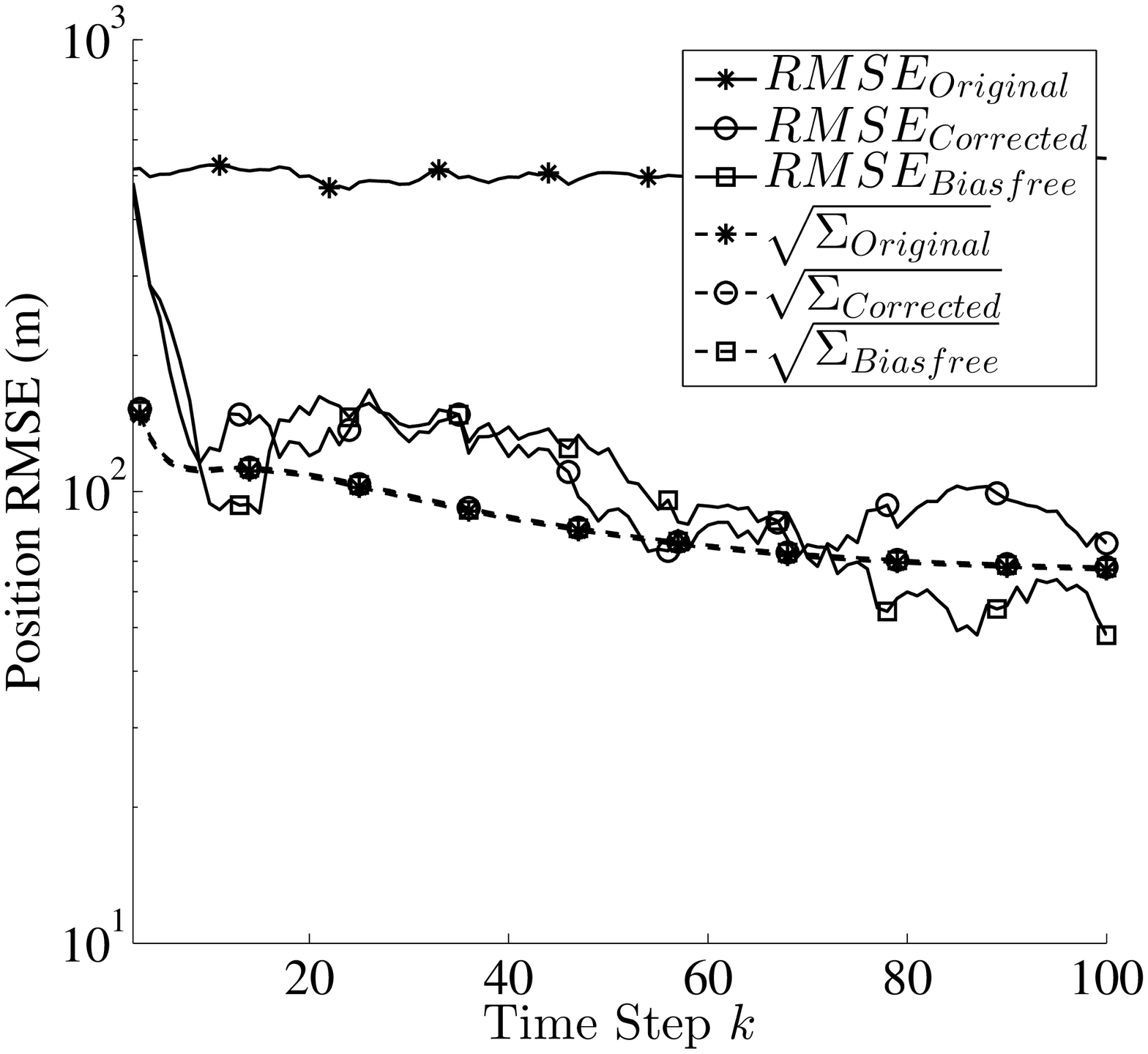}
\caption{Position RMSE \textcolor{black}{with distributed tracking} for corrected and original tracks of target 3 (set 2)}
\label{fig:paperB_Pos_set2_t3}
\end{figure}


\begin{figure}[htbp!]
\centering
\includegraphics[width=3.5in]{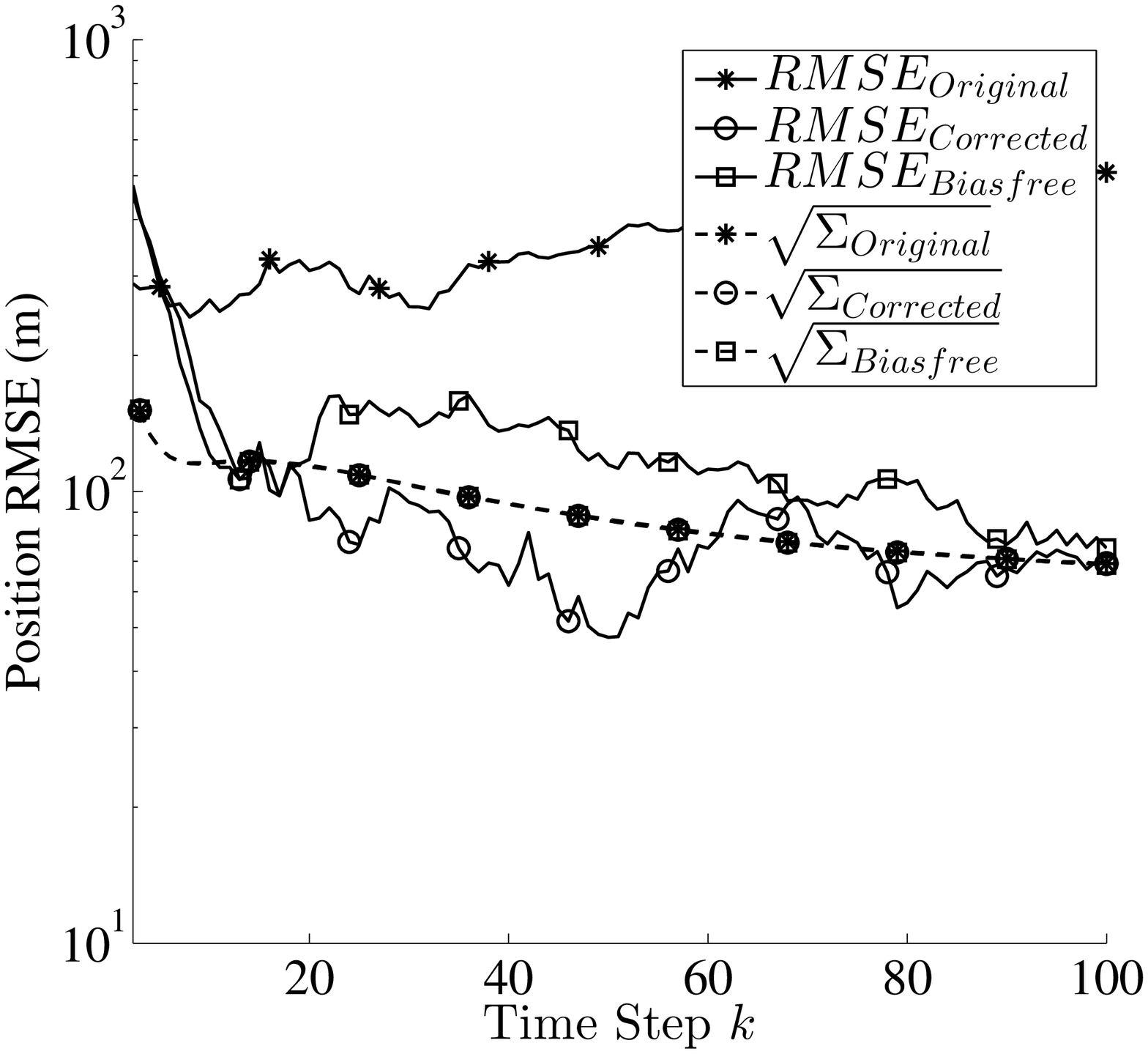}
\caption{Position RMSE \textcolor{black}{with distributed tracking} for corrected and original tracks of target 2 (set 3)}
\label{fig:paperB_Pos_set3_t2}
\end{figure}

\begin{figure}[htbp!]
\centering
\includegraphics[width=3.5in]{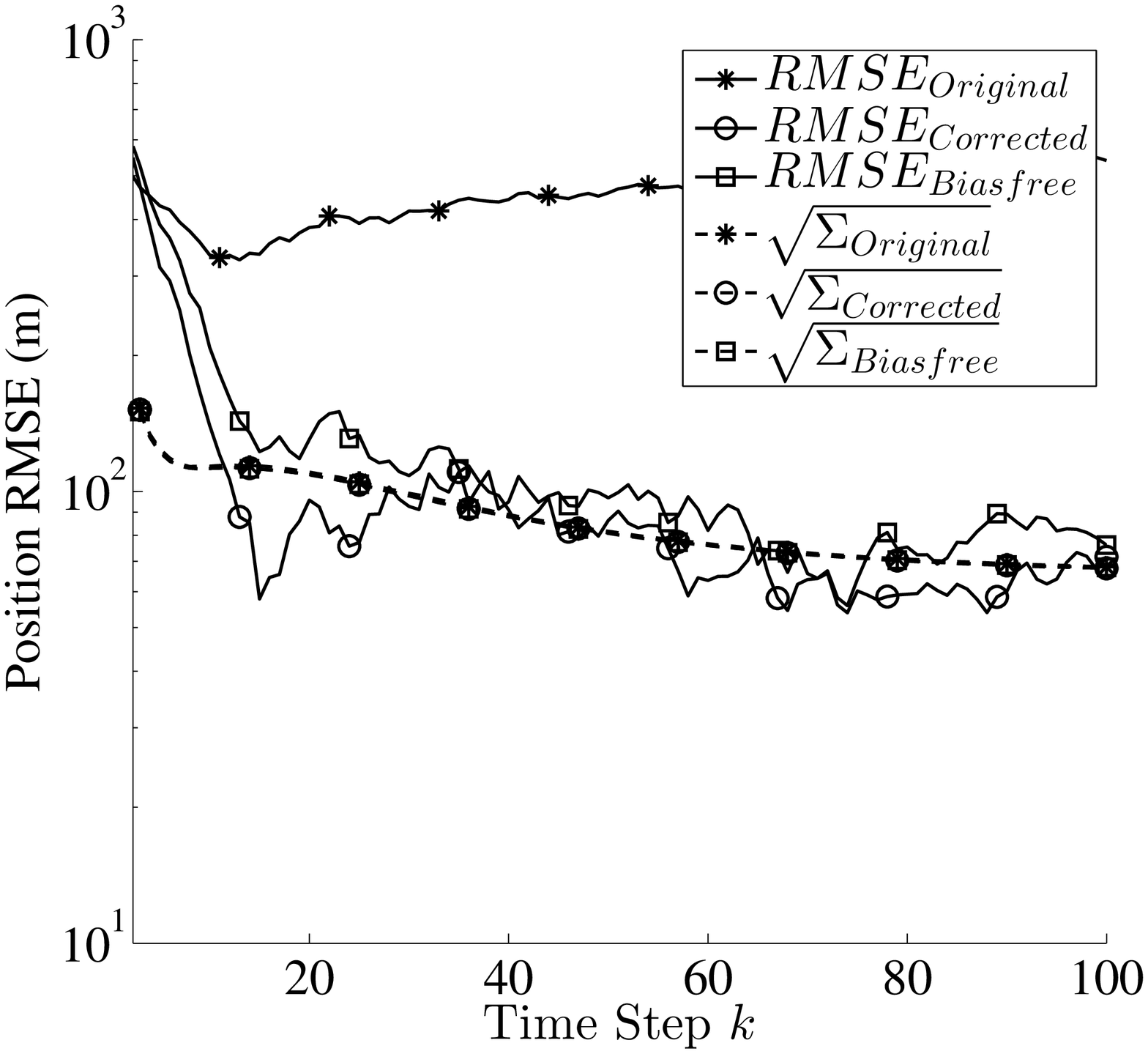}
\caption{Position RMSE \textcolor{black}{with distributed tracking} for corrected and original tracks of target 3 (set 3)}
\label{fig:paperB_Pos_set3_t3}
\end{figure}



As shown in Figures \ref{fig:paperB_Pos_set1_t2} and \ref{fig:paperB_Pos_set1_t3}, for case $1$,  Figures \ref{fig:paperB_Pos_set2_t2} and \ref{fig:paperB_Pos_set2_t3} for case $2$ and Figures \ref{fig:paperB_Pos_set3_t2} and \ref{fig:paperB_Pos_set3_t3} for case $3$, the position error is reduced significantly in terms of the position  RMSE. This demonstrates the effectiveness of the bias model proposed and the ML estimation algorithm, \ie, the GA, that is applied to the data. Although, according to the results, the correction factor varies based on the sensor--target orientation, the corrected track and its associated covariance do follow the bias--free values accurately, which demonstrates the capability of the proposed algorithm in estimating the biases. To process a batch of data with $K=100$, the computational time is $44.1 \mathrm{s}$ in MATLAB.

\subsection{\label{paperB_comparision3sensors} Bias estimation: Three--sensor \textcolor{black}{distributed} problem}
\begin{figure}[htbp!]
\centering
\includegraphics[width=3.5in]{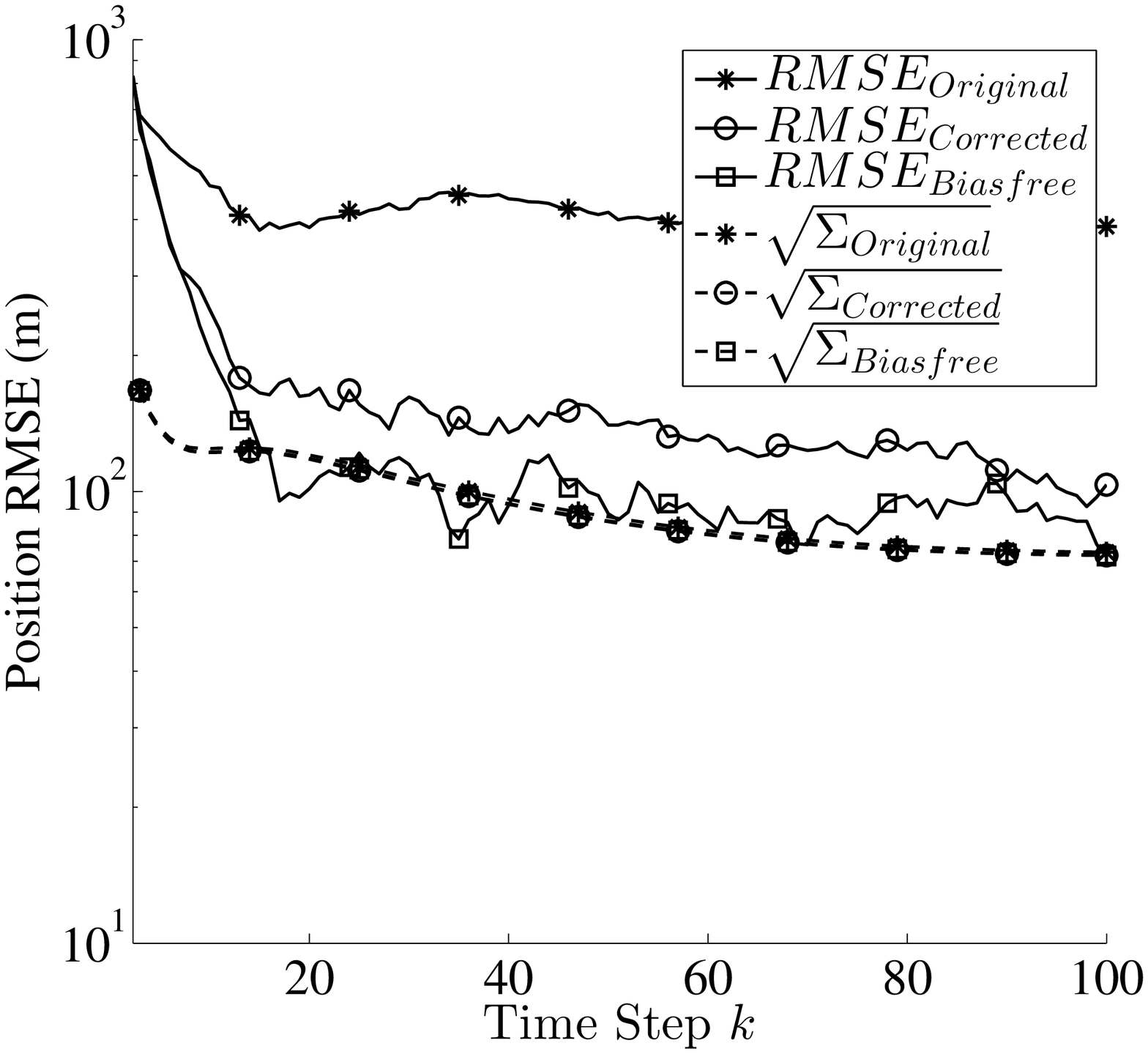}
\caption{Position RMSE of corrected and original tracks for the three--sensor distributed tracking case (Target 2)}
\label{fig:paperB_Pos_3sen_t2}
\end{figure}

\begin{figure}[htbp!]
\centering
\includegraphics[width=3.5in]{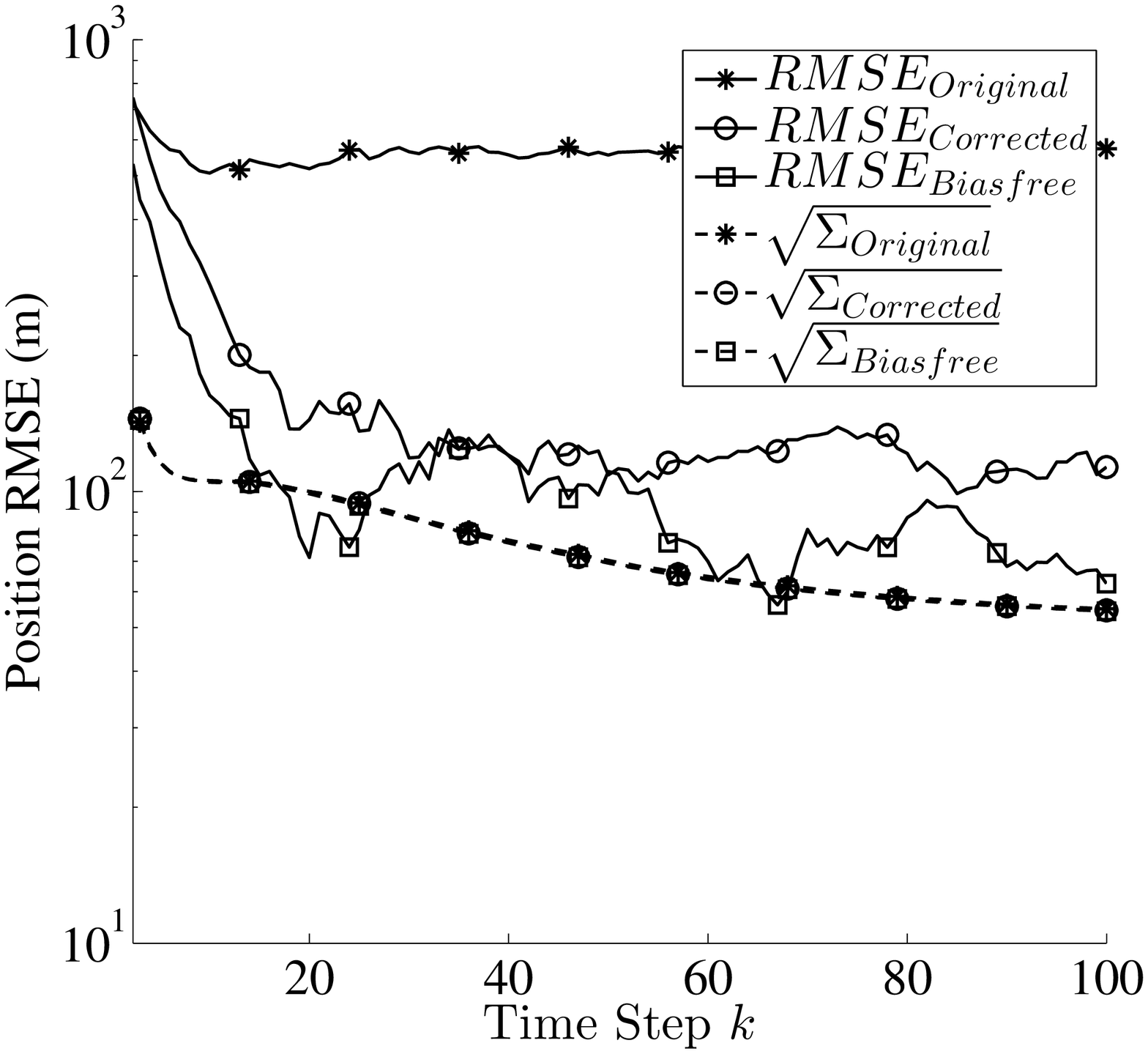}
\caption{Position RMSE of corrected and original tracks for the three--sensor distributed tracking case (Target 3)}
\label{fig:paperB_Pos_3sen_t3}
\end{figure}

To show the accuracy of bias estimation when there are only three sensors in the surveillance region, the same Genetic Algorithm is used to solve the ML problem in \eqref{paperB_eq:maxLikelihood}. The primary issue with three sensors is that one of them, here sensor $i$, is used to pair with both sensors $j$ and $m$. This leads to correlation between the two tracks over time. Assuming that the correlation is negligible, the same method is applied to estimate the biases. Further, for the case of three sensors, the bias values are set to 
\begin{eqnarray}
\mathbf{b}_{\mathrm{test}} & = & \left[\begin{array}{ccc}
0.04\mathrm{\: rad} & 0.02\mathrm{\: rad} & 0.03\mathrm{\: rad}\end{array}\right]^{\mathrm{T}}\label{paperB_eq:test3sensors}
\end{eqnarray}
to be able to distinguish between the original and corrected tracks. The constraints on the lower and upper bounds of the optimization algorithm are set to $-0.05\:\mathrm{rad}$ and $0.05\:\mathrm{rad}$, respectively. The tolerance value for the GA is also set to $1\times10^{-15}$. Figures \ref{fig:paperB_Pos_3sen_t2} and \ref{fig:paperB_Pos_3sen_t3} show the result for position estimates in Cartesian coordinates.

The corrected tracks are sill better in terms of RMSE, which means that the estimated biases are accurate enough in spite of the correlation.

\subsection{\label{paperB_realTimeGA} Real--time window--based Genetic Algorithm}
It is important for the proposed method to be able to work in real--time. For this purpose, the GA can be set to run, in each iteration, for a specific window size or duration. The settings of the GA for real time scenarios are given in Table \ref{paperB_GAWindow}.

\begin{table}
\caption{Parameter settings for real time genetic algorithm}
\label{paperB_GAWindow}
\centering{}%
\begin{tabular}{|c|c|}
\hline 
Parameter name & Value\tabularnewline
\hline 
\hline 
Lower bound & $-0.05\:\mathrm{rad}$\tabularnewline
\hline 
Upper bound & $0.05\:\mathrm{rad}$\tabularnewline
\hline 
Number of generations & $50$\tabularnewline
\hline 
Tolerance value & $1\times10^{-15}$\tabularnewline
\hline 
Window size & 10\tabularnewline
\hline 
\end{tabular}
\end{table}

\begin{figure}[htbp!]
\centering
\includegraphics[width=3.5in]{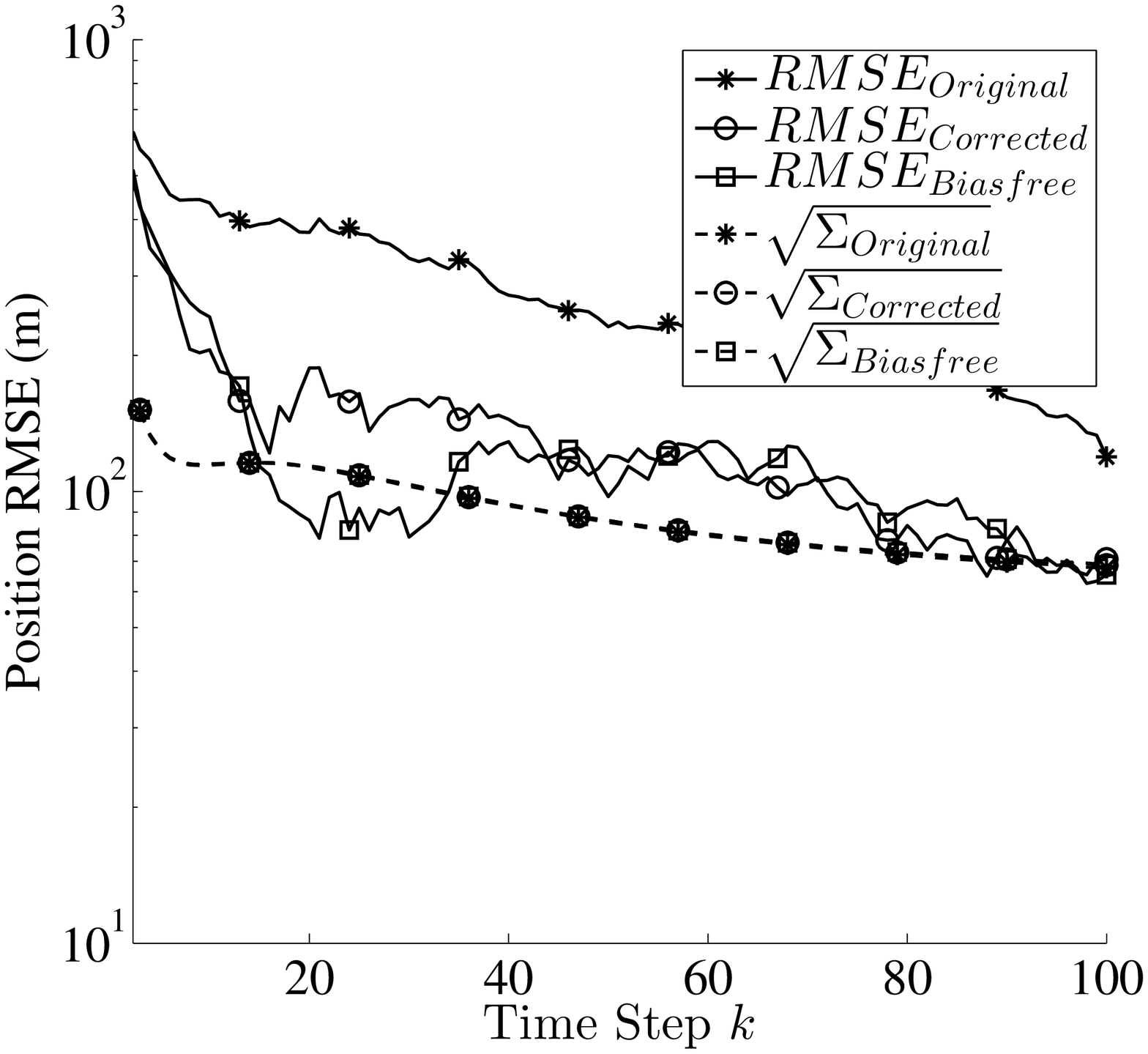}
\caption{Position RMSE for corrected and original tracks for the four--sensor distributed tracking case and window size of 10 (Target 2)}
\label{fig:paperB_Pos_10_t2}
\end{figure}

\begin{figure}[htbp!]
\centering
\includegraphics[width=3.5in]{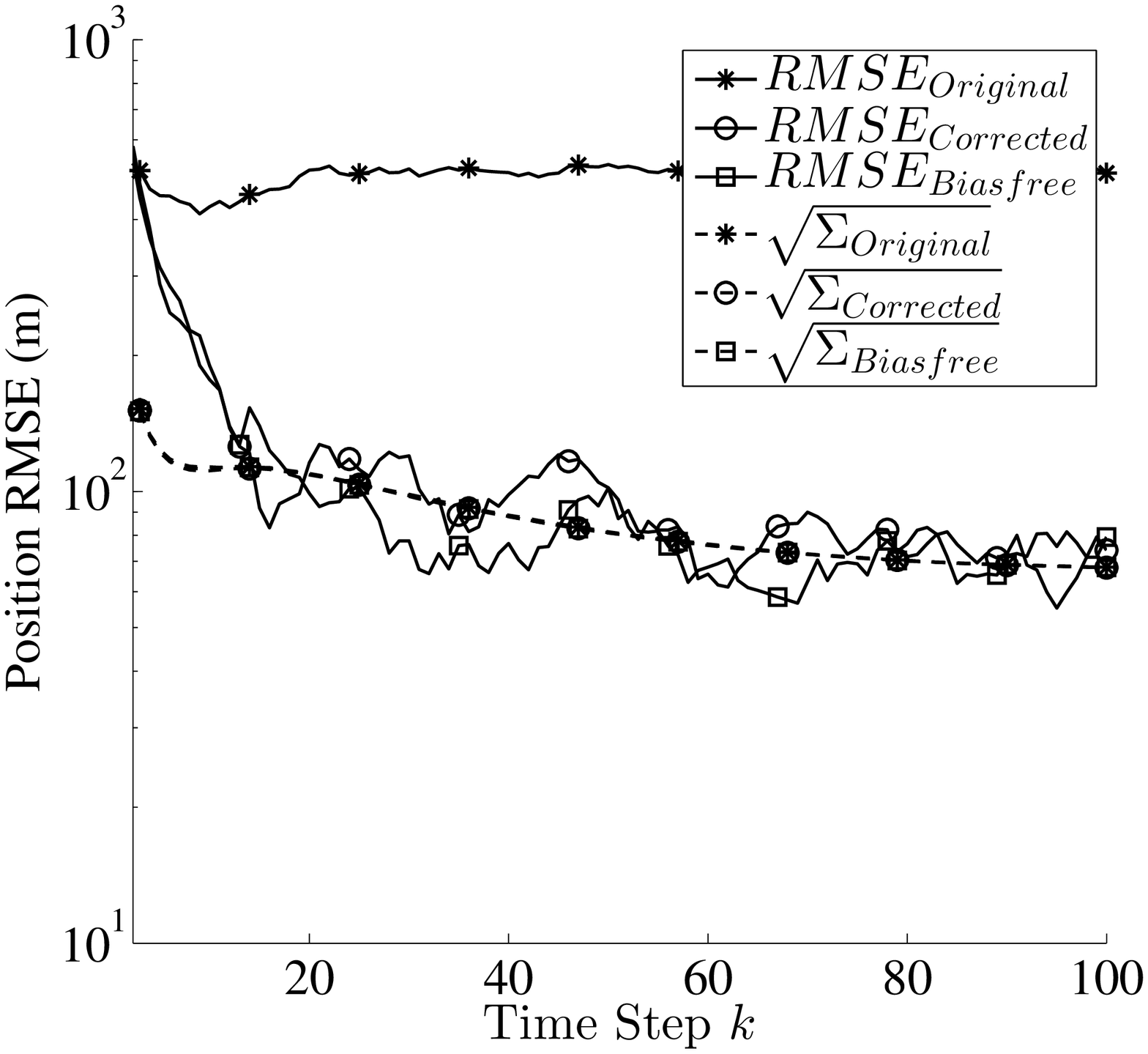}
\caption{Position RMSE for corrected and original tracks for the four--sensor distributed tracking case and window size of 10 (Target 3)}
\label{fig:paperB_Pos_10_t3}
\end{figure}

The final estimates and the population matrix in one window can be used as the initial conditions for the next window. Thus, the estimates of the biases can be used to correct the measurements at the end of processing each window of data. The results of the simulations using this approach for the case of four sensors (test 1) are given in Figures \ref{fig:paperB_Pos_10_t2} and \ref{fig:paperB_Pos_10_t3}.

As shown in Figures \ref{fig:paperB_Pos_10_t2} and \ref{fig:paperB_Pos_10_t3}, the GA is still able to find the biases with a smaller window size and fewer generations. Note that updating the biases with smaller window sizes enables the use of methods similar to \cite{taghavi2013bias,taghavi2014bias} for arbitrary number of sensors in the surveillance region.  With this setting, biases can be updated every $9.25\:\mathrm{s}$ in MATLAB. This simple example shows that even when the processing time is a crucial parameter in the design, one can still handle bias estimation in real--time using the proposed method.

\subsection{\label{paperB_falseAlarm} \textcolor{black}{Bias estimation with false measurements: Four--sensor centralized problem}}
\textcolor{black}{To demonstrate how the proposed method performs in the presence of false alarms and missed associations in a centralized fusion framework where local sensors send all their measurements instead of AMRs or bearing--only tracks, a simulation is presented in this section. The probability mass function of the number of false alarms or clutter points in surveillance volume $V$ as a function of their spatial density $\lambda$ is defined as 
\begin{eqnarray}
\mu_{\mathrm{FA}}\left(m\right) & = & e^{-\lambda V}\frac{\left(\lambda V\right)^{m}}{m!}\label{eq:paperB_falseAlarm}
\end{eqnarray}
where $m$ is the number of false alarms \cite{bar2011tracking}. With bearing--only measurements,  the volume is $V=2\pi$ ($\mathrm{rad}$). It is assumed that the average number of false alarms per unit volume in a scan, \ie, $\lambda$, is $0.5$. Also, it is assumed that $P_D=0.7$ for each sensor. Note that in this centralized case, the local sensors send all measurements (rather than AMRs or local bearing--only tracks) to the fusion node. Because the proposed method is a batch estimator and uses measurements from both sensor pairs to create a pseudo--measurement for bias estimation, false tracks are often removed prior to generating the pseudo--measurement vector, which then is sent to the bias estimator. Typically, false tracks do not exist for more than a few time steps as they are dependent on all four sensors creating false alarms at the same time steps, in the same region, and for a reasonably long interval of time.}

\textcolor{black}{To show the accuracy of bias estimation in a centralized system, the same Genetic Algorithm is used to solve the ML problem in \eqref{paperB_eq:maxLikelihood}. The RMSE values and $\sqrt{\mathrm{CRLB}\left\{ \left[\mathbf{b}\right]_{i}\right\}}$ of the ML estimates with the bias parameters as defined in \eqref{paperB_eq:test1} are shown in Table \ref{tab:4sensors_false}. Note that the CRLB values are optimistic because they do not factor in the false alarms or the missed detections and that the ML estimator does not factor in the false alarms or the missed detections explicitly. A comprehensive centralized bias estimator is under development. The focus of this paper is the decentralized one.}

\begin{table}[htbp!]
\caption{Comparison of CRLB and GA output for bias estimation of $\mathbf{b}_{\mathrm{test_{1}}}$with measurement origin uncertainty in a centralized system
($\lambda=0.5$ and $P_{D}=0.7$)}
\centering{}%
\begin{tabular}{|c|c|c|}
\hline 
Bias parameter & CRLB & GA bias estimate\tabularnewline
\hline 
\hline 
$b_{i}$ & $0.2123\times10^{-3}$ & $7.419\times10^{-3}$\tabularnewline
\hline 
$b_{j}$ & $0.2123\times10^{-3}$ & $12.85\times10^{-3}$\tabularnewline
\hline 
$b_{m}$ & $0.7674\times10^{-3}$ & $6.143\times10^{-3}$\tabularnewline
\hline 
$b_{n}$ & $0.5779\times10^{-3}$ & $4.352\times10^{-3}$\tabularnewline
\hline 
\end{tabular}
\label{tab:4sensors_false}
\end{table} 

\section{\label{paperB_conclusion} Conclusions}
\textcolor{black}{In this paper, a new mathematical model for bearing--only bias estimation in distributed tracking systems was proposed. This model was based on triangulation using the associated measurement reports or local bearing--only tracks from different sensor pairs. It was shown that the proposed bias model has the advantage of being practical in scenarios with multiple sensors. In particular, the proposed algorithm is effective when the sensor noise level and bias values are high. In addition, previously proposed algorithms were dependent on target--sensor maneuvers and/or limited to certain noise levels. The new bias model can handle any type of target--sensor motion and it is effective against $5^{\circ}$ of offset bias in each sensor and uncertainty levels up to $1.5^{\circ}$ of noise standard deviation, which is higher than what was assumed in the literature previously. Also, the proposed method can handle false alarms and missed detections in a centralized architecture. That is, the proposed algorithm is practical in scenarios with realistic sensor parameter values. Finally, a batch ML estimator was proposed to solve the bias estimation problem along with simulation results. A comprehensive centralized bias estimation algorithm with data association for bearing--only sensors is in progress.}
\ifCLASSOPTIONcaptionsoff
  \newpage
\fi



%
\bibliographystyle{Myplain}
\bibliography{references}
\begin{biography}[{\includegraphics[width=1in,height=1.25in,clip,keepaspectratio]{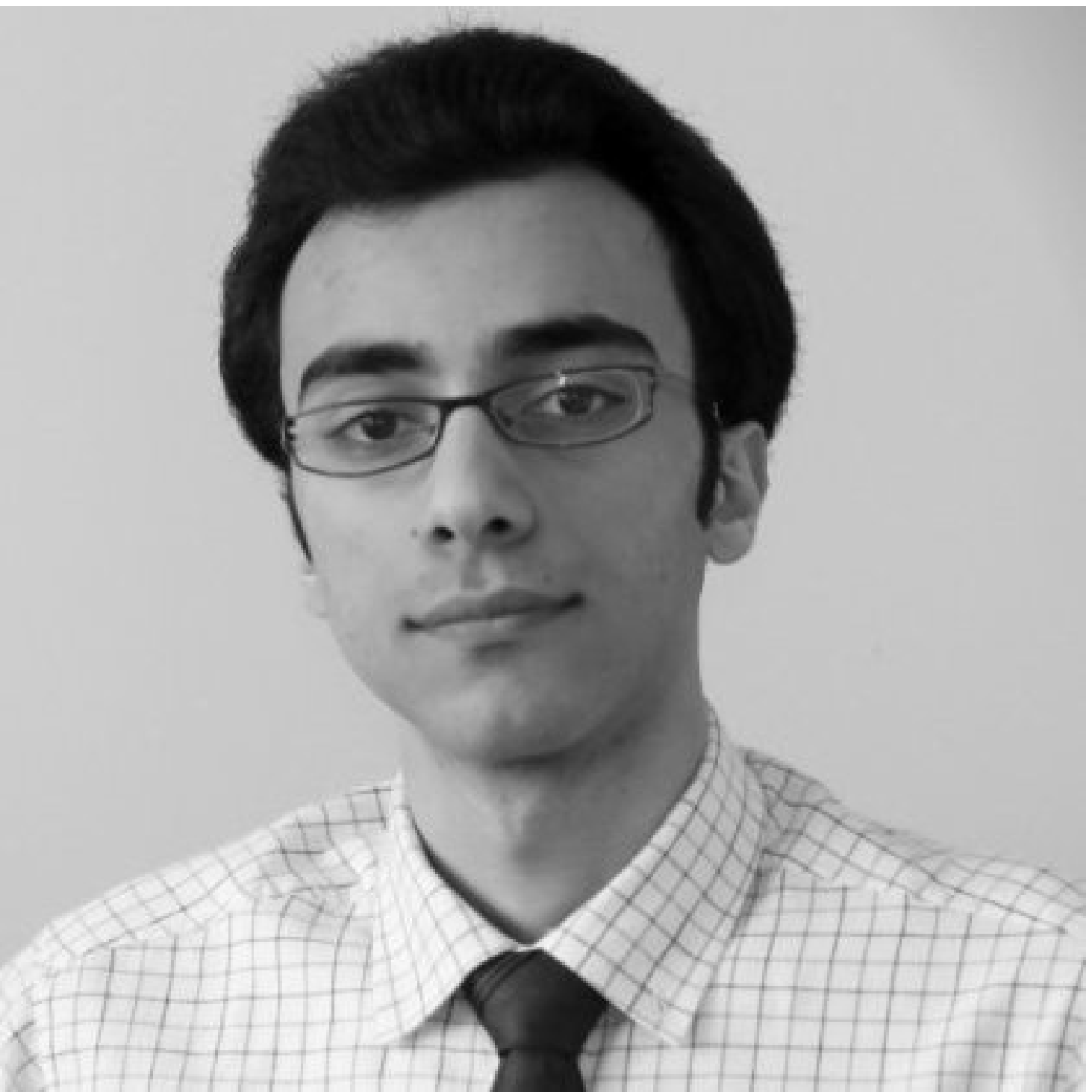}}]{Ehsan Taghavi} received the M.Sc. degree in communication engineering in 2012 from Chalmers University of Technology, Gothenburg, Sweden, where he worked on particle filter smoother. He is currently pursuing the Ph.D. degree in computational science and engineering at McMaster University, Hamilton, Canada. His research interests include estimation theory, scientific computing, signal processing, parameter estimation, mathematical modeling and algorithm design.
\end{biography}

\begin{biography}[{\includegraphics[width=1in,height=1.25in,clip,keepaspectratio]{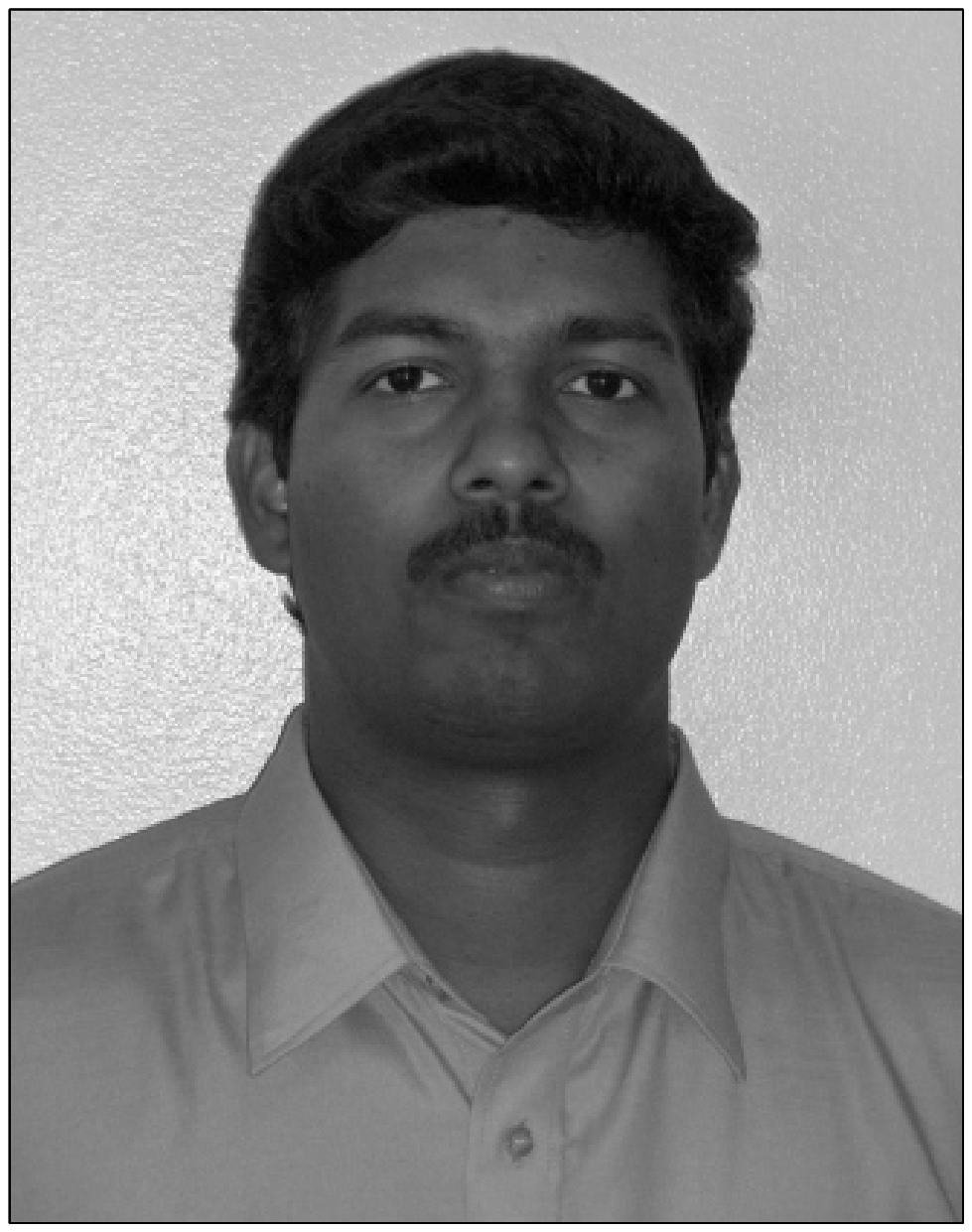}}]{Ratnasingham Tharmarasa}
was born in Sri Lanka in 1975. He received the B.Sc.Eng. degree in electronic
and telecommunication engineering from University of Moratuwa, Sri Lanka in
2001, and the M.A.Sc and Ph.D. degrees in electrical engineering from McMaster
University, Canada in 2003 and 2007, respectively.

From 2001 to 2002 he was an instructor in electronic and telecommunication
engineering at the University of Moratuwa, Sri Lanka. During 2002-2007 he was a
graduate student/research assistant in ECE department at the McMaster
University, Canada. Currently he is working as a Research Associate in the
Electrical and Computer Engineering Department at McMaster University, Canada.
His research interests include target tracking, information fusion and sensor
resource management.
\end{biography}

\begin{biography}[{\includegraphics[width=1in,height=1.25in,clip,keepaspectratio]{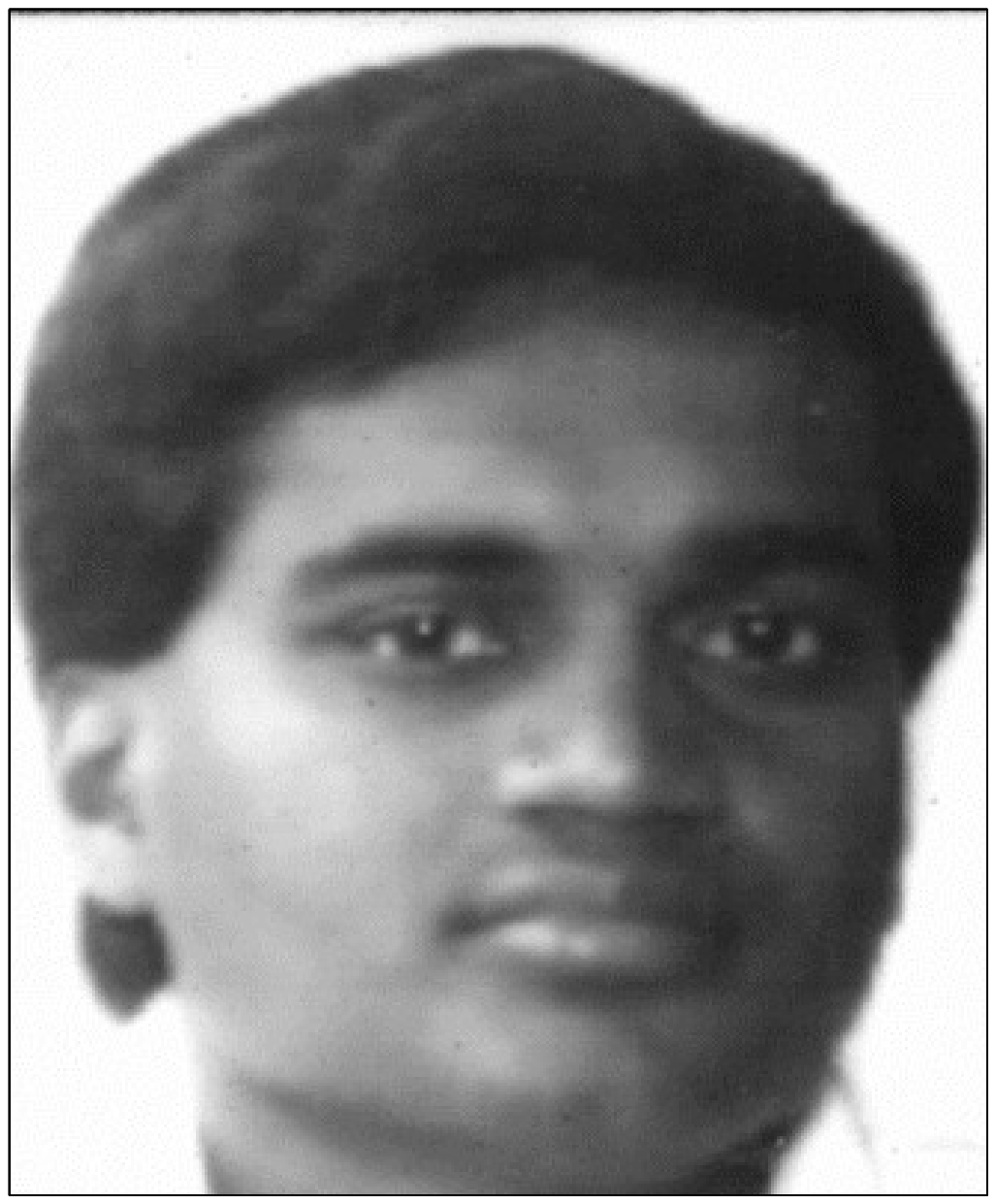}}]{Thiagalingam Kirubarajan}
(S'95–M'98–SM'03) was born in Sri Lanka in 1969. He received the B.A. and M.A.
degrees in electrical and information engineering from Cambridge University,
England, in 1991 and 1993, and the M.S. and Ph.D. degrees in electrical
engineering from the University of Connecticut, Storrs, in 1995 and 1998,
respectively.

Currently, he is a professor in the Electrical and Computer Engineering
Department at McMaster University, Hamilton, Ontario. He is also serving as an
Adjunct Assistant Professor and the Associate Director of the Estimation and
Signal Processing Research Laboratory at the University of Connecticut. His
research interests are in estimation, target tracking, multisource information
fusion, sensor resource management, signal detection and fault diagnosis. His
research activities at McMaster University and at the University of Connecticut
are supported by U.S. Missile Defense Agency, U.S. Office of Naval Research,
NASA, Qualtech Systems, Inc., Raytheon Canada Ltd. and Defense Research
Development Canada, Ottawa. In September 2001, Dr. Kirubarajan served in a
DARPA expert panel on unattended surveillance, homeland defense and
counterterrorism. He has also served as a consultant in these areas to a number
of companies, including Motorola Corporation, Northrop-Grumman Corporation,
Pacific-Sierra Research Corporation, Lockhead Martin Corporation, Qualtech
Systems, Inc., Orincon Corporation and BAE systems. He has worked on the
development of a number of engineering software programs, including BEARDAT for
target localization from bearing and frequency measurements in clutter, FUSEDAT
for fusion of multisensor data for tracking. He has also worked with Qualtech
Systems, Inc., to develop an advanced fault diagnosis engine.

Dr. Kirubarajan has published about 100 articles in areas of his research
interests, in addition to one book on estimation, tracking and navigation and
two edited volumes. He is a recipient of Ontario Premier’s Research Excellence
Award (2002).
\end{biography}

\begin{biography}[{\includegraphics[width=1in,height=1.25in,clip,keepaspectratio]{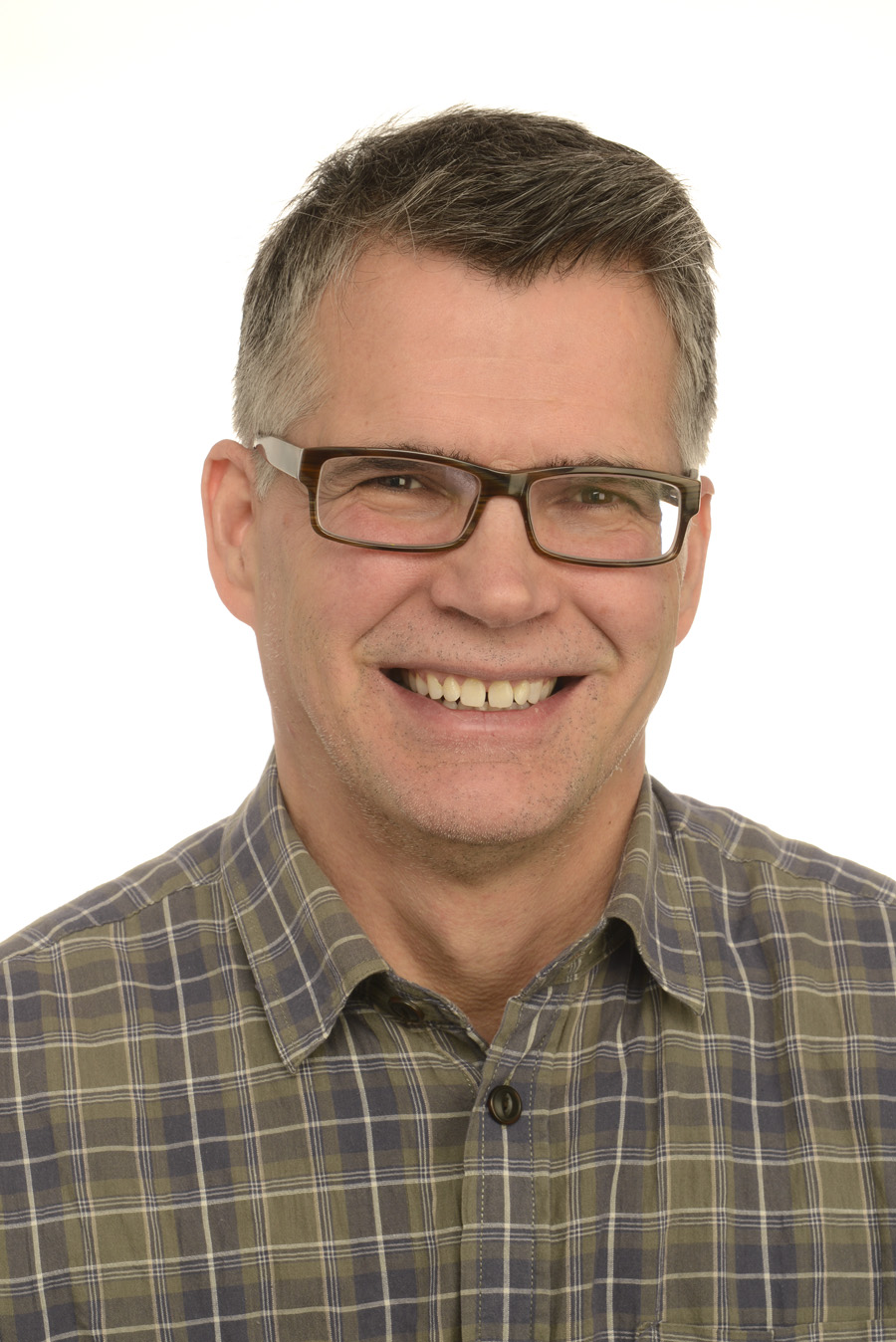}}]{Michael McDonald} received a B.Sc (Hons) degree in Applied Geophysics from Queens University in Kingston, Canada in 1986 and a M.Sc. degree in Electrical Engineering in 1990, also from Queen's University.  He received a Ph.D in Physics from the University of Western Ontario in London, Canada in 1997.  He was employed at ComDev in Cambridge, Canada from 1989 through 1992 in their space science and satellite communications departments and held a post-doctoral position in the Physics department of SUNY at Stony Brooke from 1996 through 1998 before commencing his current position as Defence Scientist in the Radar Systems section of Defence Research and Development Canada, Ottawa, Canada.  His current research interests include the application of STAP processing and nonlinear filtering to the detection of small maritime and land targets as well as the development and implementation of passive radar systems.
\end{biography}
%








\end{document}